%% file: porkchop.tex
\documentclass{sig-alternate}
\usepackage{balance,array}
\usepackage[bf]{subfigure}

\usepackage{color, mathptmx, amsmath, amssymb, amsthm, verbatim,
  mdwlist, graphicx, footnote, multirow, xspace, tikz,
  colortbl,cite,comment,inconsolata, listings}

\DeclareMathAlphabet{\mathcal}{OMS}{cmsy}{m}{n}
\let\footnotesize\small

\usepackage{algorithm}

\usepackage[noend]{algpseudocode}

\algrenewcommand{\alglinenumber}[1]{\small#1:}

\usepackage[protrusion=true,expansion=true]{microtype}
\theoremstyle{definition}
\makeatletter
\def\thm@space@setup{\thm@preskip=.5em
\thm@postskip=.5em}
\makeatother

\theoremstyle{remark}

\usepackage{scrextend}
\usepackage{breakurl}

\usepackage[hidelinks]{hyperref}

\newcommand{\minihead}[1]{{\vspace{.4em}\noindent\textbf{#1.} }}

\newcommand{\bsgd}{\textsc{ASIP-SGD}\xspace}
\newcommand{\badmm}{\textsc{ASIP-ADMM}\xspace}
\newcommand{\hogwild}{\textsc{Hogwild!}\xspace}

\newcommand{\xchg}{ASIP\xspace}
\newcommand{\poll}{\texttt{poll}\xspace}
\newcommand{\push}{\texttt{push}\xspace}

\usepackage{etoolbox}
\newcommand{\zerodisplayskips}{%
  \setlength{\abovedisplayskip}{4pt}
  \setlength{\belowdisplayskip}{4pt}
  \setlength{\abovedisplayshortskip}{4pt}
  \setlength{\belowdisplayshortskip}{4pt}}
\appto{\normalsize}{\zerodisplayskips}
\appto{\small}{\zerodisplayskips}
\appto{\footnotesize}{\zerodisplayskips}

\newenvironment{myitemize}
{
  \vspace{-.2em}
    \begin{list}{$\bullet$ }{}
        \setlength{\topsep}{0em}
        \setlength{\parskip}{0pt}
        \setlength{\partopsep}{0pt}
        \setlength{\parsep}{0pt}
        \setlength{\itemsep}{.25em}
        \setlength{\itemindent}{0em}
}
{
    \end{list}
    \vspace{-.2em}
}

\newenvironment{evalenumerate}
{

   \vspace{-.5em}
   \newcounter{qdecounter}
    \begin{list}{\arabic{qdecounter}.~}{\usecounter{qdecounter}\leftmargin=1em}
        \setlength{\topsep}{0em}
        \setlength{\parskip}{0pt}
        \setlength{\partopsep}{0pt}
        \setlength{\parsep}{0pt}
        \setlength{\itemsep}{.25em}
        \setlength{\itemindent}{0em}
}
{
    \end{list}
    \vspace{-.5em}
}

\newcommand{\argmin}{\arg\!\min}

\newcommand{\tableref}[1]{Table~\ref{#1}}
\newcommand{\figref}[1]{Figure~\ref{#1}}
\newcommand{\listref}[1]{Listing~\ref{#1}}

\newcommand{\eqnref}[1]{Eq.~(\ref{#1})}
\newcommand{\secref}[1]{Section~\ref{#1}}

\newcommand{\myalgref}[1]{Alg.~\ref{#1}}

\newcommand{\OneNorm}[1]{\| #1 \|_1}
\newcommand{\TwoNorm}[1]{\| #1 \|_2}

\DeclareMathOperator{\loss}{loss}
\DeclareMathOperator{\reg}{reg}
\newcommand{\data}{\mathcal{D}}

\def\naive{na\"{\i}ve\xspace}

\lstdefinelanguage{scala}{
  morekeywords={abstract,case,catch,class,def,%
    do,else,extends,false,final,finally,%
    for,if,implicit,import,match,mixin,%
    new,null,object,override,package,%
    private,protected,requires,return,sealed,%
    super,this,throw,trait,true,try,%
    type,val,var,while,with,yield},
  otherkeywords={=>,<-,<\%,<:,>:,\#,@},
  sensitive=true,
  morecomment=[l]{//},
  morecomment=[n]{/*}{*/},
  morestring=[b]",
  morestring=[b]',
  morestring=[b]""",
  frame=tb, captionpos=b,
  basicstyle=\small\fontfamily{pcr}\selectfont
}

\begin{document}
%
\conferenceinfo{XXX}{YYY}

\title{Asynchronous Complex Analytics\\in a Distributed Dataflow Architecture}


\author{
Joseph E. Gonzalez, Peter Bailis$^\dagger$, Michael I. Jordan, Michael J.
Franklin,\\[1mm] Joseph M. Hellerstein, Ali Ghodsi, Ion Stoica\\[1mm]
\affaddr{UC Berkeley and $\dagger$Stanford University}}

\maketitle

\input{abstract}

\input{intro}

\input{background}

\input{architecture}
\input{algorithm}
\input{implementation}

\input{evaluation}

\input{relatedwork}
\input{conclusion}

\section*{Acknowledgments}

This research was supported in part by NSF CISE Expeditions Award
CCF-1139158, LBNL Award 7076018, DARPA XData Award FA8750-12-2-0331,
the NSF Graduate Research Fellowship (grant DGE-1106400), and gifts
from Amazon Web Services, Google, SAP, The Thomas and Stacey Siebel
Foundation, Adatao, Adobe, Apple, Inc., Blue Goji, Bosch, C3Energy,
Cisco, Cray, Cloudera, EMC, Ericsson, Facebook, Guavus, Huawei,
Informatica, Intel, Microsoft, NetApp, Pivotal, Samsung, Splunk,
Virdata, VMware, and Yahoo!.

\scriptsize
\linespread{.97}
\bibliography{porkchop} \bibliographystyle{abbrv}

\input{appendix}

\end{document}

%% file: abstract.tex

\begin{abstract} 

Scalable distributed dataflow systems have recently experienced widespread adoption, with commodity dataflow engines such as Hadoop and Spark, and even commodity SQL engines routinely supporting increasingly sophisticated analytics tasks (e.g., support vector machines, logistic regression, collaborative filtering). However, these systems' synchronous (often Bulk Synchronous Parallel) dataflow execution model is at odds with an increasingly important trend in the machine learning community: the use of asynchrony via shared, mutable state (i.e., data races) in convex programming tasks, which has---in a single-node context---delivered noteworthy empirical performance gains and inspired new research into asynchronous algorithms. In this work, we attempt to bridge this gap by evaluating the use of lightweight, asynchronous state transfer within a commodity dataflow engine. Specifically, we investigate the use of asynchronous sideways information passing (ASIP) that presents single-stage parallel iterators with a Volcano-like intra-operator iterator that can be used for asynchronous information passing. We port two synchronous convex programming algorithms, stochastic gradient descent and the alternating direction method of multipliers (ADMM), to use ASIPs. We evaluate an implementation of ASIPs within on Apache Spark that exhibits considerable speedups as well as a rich set of performance trade-offs in the use of these asynchronous algorithms.
\end{abstract}

%% file: intro.tex

\section{Introduction}
\label{sec:intro}


The recent rise of large-scale distributed dataflow frameworks has enabled
widespread adoption of increasingly sophisticated analytics tasks at
scale~\cite{carey-bigdata,spark,stratosphere,mapreduce,dryad}. The
last decade has seen considerable research and industrial effort put towards
understanding how to integrate complex analytics and learning tasks
into programmer workflows~\cite{chris-feature,mlbase}, existing system
architectures~\cite{bismarck,madlib}, and new cluster compute
frameworks~\cite{mli,vowpal}.


Simultaneously, in the machine learning community, the statistical
nature of many of these analytics tasks has led to increasing interest
in exploiting \textit{asynchrony} during computation. That is, a range
of recent theoretical results has demonstrated that removing
synchronization within an emerging class of problems can yield surprising
improvements in performance. These problems can be
solved via highly concurrent update mechanisms that expose, in effect,
read-write race
conditions~\cite{dual-averaging,hogwild-coorddescent,shalev-accelerated}. As
an example, Recht et al. have demonstrated that stochastic gradient
descent---typically implemented via serializable locking (and only
proven to converge under serial execution)---can be made robust
against asynchronous processing over shared, mutable model state: in
effect, when conflicts are rare (enough), (some) staleness and data
races will not affect statistical
correctness~\cite{hogwild}. Empirically, on single-node systems, these
asynchronous algorithms have yielded order-of-magnitude improvements
in performance and are the subject of active research, even within the
database community~\cite{bismarck,dimmwitted,sgd-matrix}.


Unfortunately, these two trends stand in opposition. Architecturally,
commodity distributed dataflow systems such as Hadoop and Spark are
optimized for coarse-grained (often bulk synchronous
parallel~\cite{valiant-bsp}) data transformations and are not designed
to natively provide the fine-grained communication required for
efficient asynchronous analytics tasks. Consequently, evaluation of
these new asynchronous algorithms have been largely confined to
single-node, multi-processor (and NUMA)
context~\cite{dimmwitted,hogwild}: it is relatively unknown how the
increased latency of a distributed environment impacts their
performance and correctness guarantees. The technological trajectory
outlined by recent research suggests a divide between
widely-deployed dataflow-based cluster compute frameworks and
specialized asynchronous optimization mechanisms, which largely rely
on a shared memory abstraction~\cite{parameter-server,stale-parameter,yahoo-lda}.


In this work, we study this disconnect by addressing two key
questions.  First, can increasingly ubiquitous dataflow systems be
easily adapted to support asynchronous analytics tasks?  Second, in a
distributed dataflow environment, what are the benefits (and costs) of
these asynchronous algorithms compared with existing synchronous
implementations?  We present the design and evaluation of a simple
dataflow operator that $i.)$ enables implementation of asynchronous
complex statistical analytics (primarily, \textit{convex programming}
tasks, including Support Vector Machines and Logistic
Regression~\cite{boyd-book}) yet $ii.)$ is implementable using a
commodity dataflow engine (Apache Spark). We use this operator to
study the implications of bringing distributed asynchrony to two
classic convex programming procedures: stochastic gradient descent
(SGD)~\cite{boyd-book} and alternating direction method of multipliers
(ADMM)~\cite{boyd-admm}. This juxtaposition of traditional BSP systems
and algorithms with their incipient asynchronous counterparts yields
an opportunity to study the differences between these paradigms.


To address the first, architectural question, we codify and exploit a
common pattern in asynchronous analytics tasks. We observe that, on a
single machine, these tasks can be cast as single-stage parallel
dataflow, with shared memory acting as a communication channel between
operators. Therefore, to allow asynchronous data sharing during
distributed operation, we introduce the Asynchronous Sideways
Information Passing (ASIP) pattern, in which a set of shared-nothing,
data-parallel operators are provided access to a special communication
channel, called a \textit{ASIP iterator}, that allows
fine-grained communication across concurrent operator instances.  The
ASIP iterator abstracts the details of distribution and routing
(similar to the exchange operator; Section~\ref{sec:relatedwork}) and
allows fine-grained communication across operators as in sideways
information passing~\cite{ives-sideways}. This enables our target
convex programming routines to take advantage of asynchrony within
more general purpose distributed dataflow systems.  We present the
design and implementation of a prototype ASIP ASIP iterator
system in Apache Spark
and discuss the challenges arising from fault tolerance and
scheduling. Notably, in our implementation, the bulk of data transfer
and computation occurs via the primary iterator interface, exploiting
Apache Spark's strength of efficient parallel computation, while, in the
convex optimization routines we study, the ASIP iterator acts as a ``control plane''
for facilitating fine-grained model synchronization. 


To address the second, more algorithmic question, we evaluate the
costs and benefits of distributed, asynchronous execution within two
common analytics tasks. We first extend BSP SGD (as provided natively
in Spark via MLlib~\cite{mli}) to ASIP gradient descent, using the
ASIP iterator to ship fine-grained delta-encoded model updates
between operators (approximating a well-studied but---to our
knowledge---seldom empirically evaluated algorithm known as dual
averaging~\cite{dual-averaging}). We also extend BSP ADMM to the ASIP
setting, using the ASIP iterator to ship actual models between
parallel operators and leveraging Escrow-like divergence
control~\cite{escrow,olston-thesis} to bound drift imposed by
asynchrony.  Across a range of learning tasks, both ASIP algorithms
demonstrate speedups of up to two orders of magnitude compared to
their BSP counterparts. However, the two ASIP algorithms evince a careful
trade-off between speed and safety: the fast delta updates of ASIP GD
are remarkably efficient when data is well-behaved but can cause
instability in pathological workloads. In contrast, ASIP ADMM behaves
well across workloads but is generally slower. To the best of our
knowledge, this evaluation is the first apples-to-apples comparison of
these techniques at scale in a distributed setting and on real-world
data.

In summary, we make the following contributions:

\begin{myitemize}
\item We present a distributed dataflow operator providing
  intra-operator sideways information passing that is sufficient to
  implement asynchronous convex optimization routines within existing
  dataflow systems.

\item We present the design and implementation of two asynchronous
  convex programming routines---gradient descent and ADMM---within the
  ASIP operator, drawing on the theoretical machine learning
  literature when possible.

\item We evaluate the costs and benefits of asynchronous convex
  programming via ASIP within Apache Spark and demonstrate
  improvements in convergence rates via the use of asynchrony across a
  range of workloads.
\end{myitemize}

%% file: background.tex

\section{Background and Preliminaries}
\label{sec:background}

The increasingly common collection of ``Big Data''---or large
datasets---and use of large-scale cluster compute frameworks
have enabled the adoption of an increasingly sophisticated array of
complex analytics tasks at unprecedented scale
(Section~\ref{sec:relatedwork}). In this section, we provide brief
background on a large class of complex analytics tasks: statistical
optimization via convex programming. We outline traditional and
emerging solutions for these tasks and discuss the resulting
architectural and algorithmic tensions.

\subsection{Our Goal: Statistical Optimization}

While statistical analytics takes a variety of forms, a large class of
popular tasks can be categorized as statistical optimization
problems of the form:
\begin{equation}
\begin{aligned}
& \underset{\text{wrt. }\,  w}{\text{minimize}}
& &  f(w) := \frac{1}{|\data|}\sum_{r \in \data} \loss(w, r) + \lambda \reg(w) \label{eqn:obj} \\
& \text{subject to}
& & w \in \mathcal{C}
\end{aligned}
\end{equation}
That is for some dataset $\mathcal{D}$ (of size $n = |\mathcal{D}|$) we want to find the value of $w \in \mathbb{R}^d$ (e.g., a model) that minimizes the loss function $\loss: (\mathbb{R}^n, \data) \rightarrow \mathbb{R}$ (e.g., a measure of the error on a data point) plus a regularization function $\reg: \mathbb{R}^n \rightarrow \mathbb{R}$ that penalizes complex models.
The parameter $\lambda$ controls the tradeoff between accurately
fitting the data and overfitting the model to the provided data.
In general, $w$ may be constrained to lie in some convex set
$\mathcal{C}$ (e.g., the set of feasible combinations of guns and butter).  In this work we will focus the unconstrained setting $\mathcal{C} = \mathbb{R}^d$.

For example, in portfolio optimization, the loss function might
encode the expected (negative) payoff of a stock portfolio $w$, the regularization function might encode some measure of risk, and the
constraints could correspond to a limited budget.
Alternatively, in many machine learning and analytics applications,
the loss function encodes the prediction error (e.g., squared error) of a model
$w$ according to a training set $\mathcal{D}$, the $\reg$ function prevents
overfitting, and the constraints $\mathcal{C}$ may impose positivity
on the weights to preserve some notion of end-user model interpretability.

In this work, we specifically focus on \textit{convex programming}
problems---that is, problems for which the solution space is shaped
such that any local minimum is also a global minimum. This convexity
allows us to make strong guarantees about the theoretical behavior of
tasks such as logistic regression, support vector machines, and
portfolio optimization. Convex optimization represents a well-studied
research area in the mathematical optimization community
(Boyd\cite{boyd-book} provides a good technical reference). While many
useful statistical analytics tasks---like deep learning, factor model
fitting, and LMF recommendation---are \textit{non}-convex, solvers for
convex programming tasks like those we study here are often used to
achieve approximations for these
problems~\cite{hogwild,bismarck,brain}.

\subsection{Convex Programming and Asynchrony}

The convex optimization literature contains numerous
techniques and associated theoretical analyses for solving convex programs.
Here, we present intuition behind one common convex program solver and
common and emerging strategies for parallelizing it.

As a guiding example, consider the \textit{gradient descent}
algorithm. Gradient descent iteratively moves through the solution
space by repeatedly evaluating the rate of change, or gradient, at the
current solution and greedily taking steps towards solutions in the
``steepest'' direction (measured by calculating the derivative of the
objective function). For readers familiar with hill climbing
algorithms, gradient descent is effectively hill climbing (descending)
while changing all variables at once according the direction of
steepest descent. A popular variant of gradient descent is
\textit{stochastic gradient descent} (SGD), which evaluates the
gradient by only looking at a random subset of the data before taking
a step, in effect reducing the total number of scans over the data.

\minihead{Theory: Serial algorithms} In the theoretical literature
(and in the simplest case), gradient descent and SGD are typically
expressed as serial algorithms. This is because the gradient
calculation at each step is immediately dependent on the calculation from the previous step. Moreover, a serial execution also simplifies analysis
of convergence and runtime.

\minihead{BSP: Batched intermediate computation} If we wish to use an
algorithm like SGD on large datasets, it would be advantageous to
parallelize the calculation. A classic technique---adopted by
large-scale machine learning frameworks such as
MLbase/MLlib~\cite{mlbase}---is mini-batch gradient descent in which,
at each step, the average gradient is calculated in parallel for a
random subset of the data and then applied to the previous best
solution.
This is, in effect, BSP execution,
with the gradient evaluation done in parallel and the actual gradient
step performed serially.

\minihead{Asynchrony: Breaking the barrier} Recent work in the machine
learning community has examined an alternative approach: instead of
simply parallelizing gradient computation, a set of parallel solvers
proceeds entirely concurrently, relying on \textit{asynchronous}
communication of intermediate gradient steps. As an example, the
\hogwild implementation of SGD places a single copy of the model in
the memory of a multi-core server and runs multiple worker processes
that simultaneously run gradient steps in parallel~\cite{hogwild}. Each gradient
update may partially or completely overwrite or be overwritten by another solver's update,
leading to non-serializable execution. However, empirically, this
``lock-free'' approach can deliver substantial speedups
over serializable mechanisms like locking. Dropping synchronization
barriers improves performance without compromising correctness (at
least on a single machine).

The rationale behind \hogwild is that, with sufficiently rare
conflicts and sufficiently fast communication (i.e., cache coherency
delays on the order of tens or hundreds of cycles), correctness can
still be guaranteed---even theoretically. The statistical robustness
inherent in the SGD algorithm as well as limited read/write
inconsistency still leads to a good solution---without incurring the
overhead of more coordination-intensive approaches like BSP or locking.

This asynchronous approach has recently been applied to diverse
problems including coordinate descent, deep learning, and portfolio
optimization~\cite{bismarck,dimmwitted,shalev-accelerated,hogwild-coorddescent,dual-averaging}. A
burgeoning cottage industry of machine learning researchers has begun
to extend asynchronous execution strategies to an increasing number of
optimization tasks.

\subsection{Asynchrony and Big Data Systems}

While the benefits offered by asynchrony are compelling, they are
currently at odds with the dominant class of large-scale data
processing systems. The architectural underpinnings of systems such as
Spark, Hadoop, and Tez favor large-scale, bulk data movement and
transformation via shared-nothing parallel dataflow. The fine-grained
communication required by this new class of asynchronous algorithms is
largely unsupported by these system architectures and
implementations.

In response, the machine learning community has begun to explore
alternative abstractions and systems to leverage asynchrony. For
example, in a many-core NUMA setting, DimmWitted~\cite{dimmwitted}
exposes a range of options for asynchronous data sharing, including
both shared-nothing and shared-model communication and outperforms
general cluster compute frameworks like Spark by orders of
magnitude. In a distributed context, we similarly observe a shift
towards specialized solutions: several recent proposals resemble
distributed shared memory---in effect, specialized key-value stores
that serve as a point of rendezvous for parallel worker tasks
(Section~\ref{sec:relatedwork}).

In this paper, we question this need to abandon general-purpose
parallel frameworks like Apache Spark and Hadoop when performing asynchronous statistical optimization \textit{in the distributed
  setting}. Specifically, to retain the strengths, considerable
engineering investments, and, pragmatically, large (and growing)
install bases of these frameworks we develop a method to enable
asynchronous convex programming \emph{within} a general-purpose
distributed dataflow framework.  Somewhat surprisingly, enabling these
optimization tasks only requires a minor change to their popular
BSP-like iterator interface. However, this in turn raises several
new challenges, including support for fault tolerance and efficient task scheduling.

In the process of evaluating our implementation, we provide the first
direct evaluation of several asynchronous convex programming techniques
in a single distributed environment. 
While there has been strong interest in the multi-core setting, with excellent experimental evaluations such as~\cite{dimmwitted}, there is a currently lack of understanding of 
the benefits of asynchrony in the distributed setting, which is often
less well-behaved than the single-node context under which the
assumptions in the theoretical analysis are more likely to apply. For example, we are not aware of a comparison between, say,
Spark or Hadoop, and asynchronous SGD on more than one server. With few
exceptions, it is largely unknown what benefits (or costs)
asynchronous tasks will bring in a practical distributed
environment. This task is more algorithmic in nature but is crucial to
understanding the utility of the systems challenges we address.






%% file: architecture.tex

\section{The ASIP Programming Model}
\label{sec:architecture}

In this section, we introduce the ASIP programming model---a modest
modification to sideways information passing that reconciles
asynchronous convex programming tasks and iterator-based dataflow
programs. Using the interface from this section, we will implement
ASIP algorithms for two (previously) synchronous distributed convex
programs in the next section and defer discussion of architectural and
implementation details until Section~\ref{sec:implementation}.

\subsection{A Common Programming Pattern}


To begin, we observe that asynchronous optimization routines
frequently embody a common pattern: a set of otherwise shared-nothing
solvers operate in parallel while periodically sharing progress via a
common communication channel. In a single-machine setting, this point
of rendezvous is most often a shared weight vector, with the
communication being performed by the CPU cache coherence protocol. For
example, in \hogwild SGD, parallel solvers process separate partitions
of input (e.g., training examples), repeatedly and independently of
one another, save for their model state---which is shared across all
solvers.

\begin{figure}
\includegraphics[width=\linewidth]{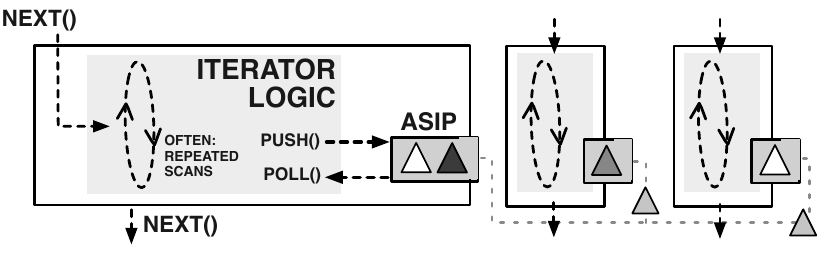}\vspace{-1em}
\caption{ASIP Iterator Architecture. A set of shared-nothing dataflow operators
  communicate asynchronously via a shared ASIP iterator, allowing
  efficient and concise implementation of complex analytics tasks
  without compromising the core dataflow abstraction.}
\label{fig:architecture}
\end{figure}

In a dataflow-based programming model, this shared state presents an
obstacle to distributing these asynchronous solvers. Unlike a single
machine, a distributed dataflow engine has no natural point of
rendezvous for asynchronous, in-band solver communication. We opt for
an iterator-based approach in order to preserve compatibility with
existing dataflow-based runtimes and programs. This explicit control
over remote operations makes their cost explicit within the
programming model~\cite{graefe-xchg}: in the distributed environment,
communication is expensive, and the desirable properties of shared
memory programming as in \hogwild are negated. Specifically, latency
(and therefore staleness) is higher, data exchange is substantially
more expensive due to serialization and networking overheads, and
communication no longer comes for ``free,'' (i.e., provided by the
underlying hardware~\cite{waldo,dsm-book}). In this section, we
describe a simple modification to sideways information passing that
allows us to make this distributed asynchronous iteration possible.



\subsection{The ASIP Interface}

The Asynchronous Sideways Information Passing (ASIP) pattern provides
a single-operator dataflow interface offering asynchronous
intra-operator communication via fine-grained iterator-based message
transfer (Figure~\ref{fig:architecture}). Specifically, ASIP augments
the standard dataflow iterator interface (requiring the implementation
of a pull-based \texttt{next()} method) with a special shared
communication channel---the \textit{ASIP}
iterator~\cite{exchange} (\xchg).\footnote{In contrast with
  traditional dataflow, where data is pulled from the ``bottom'' to
  the ``top'' of the dataflow graph, the ASIP iterator allows
  both push and pull data transfers from ``side'' to ``side''---that
  is, perpendicular to primary data flow. This is similar in spirit to
  Sideways Information Passing (Section~\ref{sec:relatedwork}) and is
  captured by a simple and familiar interface: the iterator.} Each
implementation of a ASIP iterator can use this \xchg iterator to
communicate with other instances of the same ASIP iterator running in
the same stage of the physical dataflow graph. Thus, as the name
suggests, ASIP is a natural extension of sideways information passing
(SIP)~\cite{ives-sideways}, albeit applied to asynchronous
communication.

In contrast with traditional dataflow models, under which operator
outputs are routed to successive stages in the dataflow execution
graph, ASIP outputs are routed to other instances of physical
operators within the \textit{same} stage, ASIP to typical
dataflow.\footnote{It is conceivable that ASIP outputs might be routed
  towards multiple, different operators. However, in our study of
  convex programming applications here, we did not encounter such a
  requirement, which simplified much of the design of our prototype
  ASIP implementation. We view such extensions as worthwhile future
  work.} Like the exchange operator~\cite{exchange}, the exact details
of cross-operator distribution and physical layout are opaque to the
operator implementation~\cite{graefe-xchg}. Instead, programmers using
ASIP implementers can simply treat the ASIP iterator as a
special operator from which they can send and receive intermediate
data from concurrently-executing physical operator instances within
the same datflow stage.

\begin{figure}

\subfigure[Traditional BSP]{
\includegraphics[width=1.9in]{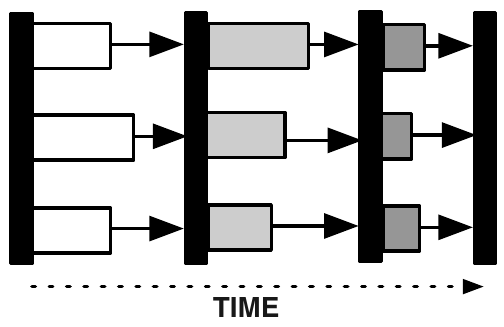}
\label{fig:bsp-comm}
}
\hfill
\subfigure[ASIP]{
\includegraphics[width=.94in]{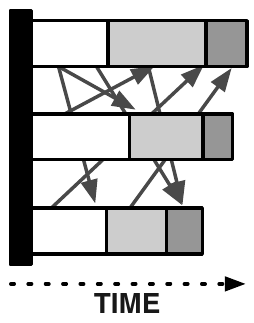}
\label{fig:ASIP-comm}
}\vspace{-1em}

\caption{Communication patterns for BSP and ASIP. Time runs from left
  to right, with barriers illustrated as vertical black bars and local
  computation as colored blocks.}
\label{fig:bsp-vs-ASIP}
\end{figure}

Unlike a BSP model, in which communication between operators is
delayed until the end of each process's pass through its partition
data, under ASIP, processes communicate asynchronously via \xchg. Thus,
ASIP adapts the BSP model by eliminating the explicit synchronous
barrier in each step of computation and instead allows individual
steps to proceed out of phase (Figure~\ref{fig:bsp-vs-ASIP}).

For now, we restrict our attention to simple, best-effort, uniform
broadcast communication (see
Section~\ref{sec:implementation}). However, like exchange and other
uses of SIPs, the actual instantiation and configuration of the
ASIP iterator can be controlled by the run-time system or, if
desirable, by user code. Notably, since \texttt{poll} does not support
blocking, a ASIP iterator instance need not actually run in parallel
in order to produce output. As we discuss in
Section~\ref{sec:relatedwork}, this interface also resembles a
restricted version of a Fjord~\cite{fjord}.

We present the actual ASIP interface in Table~\ref{table:interface}. In
addition to a standard input iterator (\texttt{child}) supporting a
standard \texttt{next()} call, a ASIP operator implementation requires
a second input, \xchg, representing the ASIP iterator. Using
\xchg, a ASIP iterator instance can both perform non-blocking reads
(via \texttt{poll}) and send messages (via \texttt{push}) to other
physical instances of the same iterator.

To demonstrate the use of the ASIP interface for asynchronous convex
programming tasks, in the next section, we port two parallel BSP
learning tasks to the ASIP model. We will discuss the actual
implementation of an ASIP, including distribution, message delivery,
and fault-tolerance mechanisms in Section~\ref{sec:implementation}.

\begin{table}[t!]
\begin{center}
\begin{tabular}{|l|l|}
\hline
\textbf{ASIP Operator Inputs} & \textbf{Operational Description}\\\hline
Input: \texttt{child} & Access to preceding iterator\\
\hspace{2.65em}\texttt{.next()}: $record$ & Fetch next tuple\\\hline
Input: \xchg & Intra-operator communication\\
\hspace{2.65em}\texttt{.push(m\textrm{: }$msg$)} & Send message to others\\
\hspace{2.65em}\texttt{.poll()}: $msg$ & Non-blocking receive\\\hline
\end{tabular}
\end{center}\vspace{-1em}
\caption{ASIP Programming Model. Each ASIP operator provides an
  implementation of the \texttt{next()} interface using the listed
  inputs above: a standard \texttt{input} iterator and a \xchg
  iterator for intra-operator communication.}
\label{table:interface}
\end{table}




%% file: algorithm.tex

\section{Programming with ASIP}
\label{sec:algorithm}

With the ASIP interface in hand, we now demonstrate how to port two
popular BSP convex programming algorithms to ASIP. Our goal is to
demonstrate how ASIP can facilitate asynchronous analytics,
and \textit{not} to innovate on new machine learning
algorithms. In this section, we present intuition and pseudocode for
each. For interested readers, we provide a more mathematically
rigorous treatment in the appendix.

\subsection{ASIP Stochastic Gradient Descent}
As we discussed in Section~\ref{sec:background}, stochastic gradient
descent is a popular algorithm for convex programming due to both its
relative simplicity and robust behavior in practice. Moreover, the
literature already provides evidence of the power of asynchronous SGD
in the single-machine case~\cite{dimmwitted,hogwild}. Therefore, SGD
makes an excellent candidate for evaluation in a distributed dataflow
system.

We begin with the traditional, BSP-based SGD and remove its synchronous
barrier. Recall that in BSP-based SGD, we compute a gradient for a
number of samples in parallel, then synchronously collect and sum the gradients before
taking a step. In ASIP-based SGD (\bsgd), we avoid this blocking and
instead allow individual parallel workers to proceed in
parallel---conceptually, just like \hogwild, albeit with explicit
state transfer between workers.

\begin{algorithm}[t!]
\small
\caption{\bsgd}
\label{alg:rapl}
\newcommand{\myindent}{\hspace{-1em}}
\begin{addmargin}[-1em]{0em}

\begin{algorithmic}[1]
\Statex{\textbf{\textit{ASIP Input}}}
\Statex $input$: random access to partition's training data, \xchg:
iterator queue, $commRate$: minimum communication time

\Statex{\textbf{\textit{ASIP Iterator}}}\vspace{.25em}

\State $\mathbf{w} \in \mathbb{R}^d$: model $\gets \mathbf{0} $
\State $t \in \mathbb{I}$: integer $\gets 1$
\For{$t \gets 1 \dots T$}
  \While{$\Delta \gets \xchg\texttt{.poll()}$ is not null}
    \State $\mathbf{w} \gets \mathbf{w} - \frac{\eta}{\sqrt{t}}\Delta$
  \EndWhile
  \State $(\mathbf{x},label) \gets$ random element of $input$
  \State $\Delta \gets \nabla_w \loss(w, (x,label)) + \lambda \nabla_w\reg(w) $
  \State $\mathbf{w} \gets \mathbf{w} - \frac{\eta}{\sqrt{t}} \Delta$
  \If{$commRate$ has elapsed}
    \State $\xchg\texttt{.push(}\Delta\texttt{)}$
  \EndIf
\EndFor
\State{\Return{$w$}}
\end{algorithmic}
\end{addmargin}
\end{algorithm}

To facilitate information transfer between workers without blocking,
\bsgd makes use the ASIP \xchg iterator. A direct ASIP interpretation of
\hogwild would simply \push each update onto \xchg and \poll on \xchg
for new updates, blindly applying them to local model state---in
effect, approximating read-write shared memory that is replicated with
each worker. However, this is potentially unwise: in addition to
serialization overheads, the latency of distributed \xchg messaging is
orders of magnitude (hundreds to tens of thousands of times) slower
than cache coherency protocols, increasing the chance of solver
divergence in the event of conflicting updates. Instead, \bsgd workers
\push their \textit{gradients} to \xchg and, upon receipt of a new
gradient (via \poll), apply the gradient (i.e., add it) to their
current model.

In the appendix, we compare this algorithm with a recently proposed
distributed optimization technique (with well-understood
theoretical---if not empirically-studied---properties) called dual
averaging~\cite{dual-averaging}.  In practice, we found that the
ASIP-SGD algorithm described here performs slightly better (and is
easier to explain).  Our ASIP implementation of SGD requires models to
be fully replicated, but we discuss partial replication in the next
section.

\subsection{ASIP ADMM}
\label{sec:admm}

\begin{algorithm}[t!]
\small
\caption{\badmm}
\label{alg:badmm}
\newcommand{\myindent}{\hspace{-1em}}
\begin{addmargin}[-1em]{0em}

\begin{algorithmic}[1]
\Statex{\textbf{\textit{ASIP Input}}}
\Statex $input$: random access to partition's training data, \xchg:
iterator queue, $commRate$: minimum communication time

\Statex{\textbf{\textit{ASIP Iterator}}}\vspace{.25em}

\State $\mathbf{w} \in \mathbb{R}^d$: primal $\gets \mathbf{0} $
\State $\mathbf{\mu} \in \mathbb{R}^d$: dual $\gets \mathbf{0} $
\State $\mathbf{\bar{w}} \in \mathbb{R}^d$: primal avg. $\gets \mathbf{0} $
\State $\mathbf{\bar{\mu}} \in \mathbb{R}^d$: dual avg. $\gets \mathbf{0} $
\State $\mathbf{z} \in \mathbb{R}^d$: consensus $\gets \mathbf{0} $

\For{$k \gets 1 \dots K$}
  \State $t \in \mathbb{I}$: integer $\gets 1$
  \While{change in $\mathbf{w} > \epsilon$}
    \Comment Solve the primal
    \State $(\mathbf{x},label) \gets$ random element of $input$
    \State $\mathbf{w} \gets \mathbf{w} - \frac{\eta}{\sqrt{t}} \left( \nabla_w \loss(\mathbf{w}, (\mathbf{x},label)) + \mathbf{\mu} + \rho \left(\mathbf{w} - \mathbf{z}\right)\right)$
  \EndWhile
  \If{$commRate$ has elapsed}
    \State $\xchg\texttt{.push(}(\mathbf{w} - \mathbf{w}_\text{old}, \mathbf{\mu} - \mathbf{\mu}_\text{old})\texttt{)}$
    \State $\mathbf{w}_\text{old} \gets \mathbf{w}$; $\mathbf{\mu}_\text{old} \gets \mathbf{\mu}$
  \EndIf
   \While{$(\Delta_{\bar{w}}, \Delta_{\bar{\mu}}) \gets \xchg\texttt{.poll()}$ is not null}
    \State $\mathbf{\bar{w}} \gets \mathbf{\bar{w}} - \Delta_{\bar{w}}$
    \State $\mathbf{\bar{\mu}} \gets \mathbf{\bar{\mu}} - \Delta_{\bar{\mu}}$
  \EndWhile
  \State $\mathbf{z} \leftarrow \argmin_{\mathbf{z}} \, \lambda  \reg(\mathbf{z}) + \frac{\mathbf{z}^T p \rho}{2} \left(\mathbf{z} - 2 \mathbf{\bar{w}} - \mathbf{\bar{\mu}}\right)$
  \Comment Consensus update
  \State $\mathbf{\mu} \gets \mathbf{\mu} + \rho \left(\mathbf{w} - \mathbf{z}\right)$
  \Comment Update the dual
\EndFor
\State{\Return{$\bar{w}$}}
\end{algorithmic}
\end{addmargin}
\end{algorithm}

As a second mechanism for study, we examine the Alternating Direction
Method of Multipliers, or ADMM, a popular synchronous and distributed
convex programming routine from the optimization
literature~\cite{boyd-admm}. Given an objective function of the form
in \eqnref{eqn:obj}, ADMM repeatedly invokes (via BSP) a partitioned
set of local solvers (say, a single-site implementation of SGD) with a
modified, decomposable version of the original convex programming
problem. By carefully manipulating this objective, ADMM iteratively
guides each local solver towards an consistent global optimum.

\minihead{Theory} We can rewrite the original convex programming problem \eqnref{eqn:obj} as the equivalent problem:
\begin{equation}
\begin{aligned}
& \underset{\text{wrt. } w_1, \ldots, w_p, z}{\text{minimize}}
& &  \lambda \reg(z) + \frac{1}{|\data|}\sum_{i = 1}^p\sum_{r \in \data_i} \loss(w_i, r)  + \frac{\rho}{2}\TwoNorm{w_i - z}^2\label{eqn:admmobj} \\
& \text{subject to}
& & \forall_i: w_i = z
\end{aligned}
\end{equation}
where we have introduced a separate variable $w_i \in \mathbb{R}^d$ for each of the $p$ machines, a consensus variable $z \in \mathbb{R}^d$, and required that all variables have exactly the same value $w_i = z$.
As a consequence, the additional quadratic penalty term $\frac{\rho}{2}\TwoNorm{w_i - z}^2 = 0 $ and therefore plays no role in the final answer.
However, this seemingly superfluous transformation accomplishes two important goals.
First, the introduction of a separate variable $w_i$ for each machine allows us to apply Lagrangian dual techniques to alternate between solving each sub-problem in isolation and then adjusting the consensus value $z$ and a dual penalty term to ensure that the independent solutions eventually agree.
Second, the additional quadratic penalty term smooths each of the sub-problems by the constant $\rho$ enabling efficient subproblem solvers and stabilizing the solution.

The resulting algorithm can be neatly cast into the two phases of the BSP model. In the first phase, each machine operates in isolation by updating the dual variables $\mu^k$ based on the previous consensus value:
\begin{equation}
\mu_i^{k+1} \leftarrow \mu_i^k + \rho (w_i^{k} - z^{k}) \label{eqn:dual}
\end{equation}
and then resolving the augmented sub-problem:
\begin{equation}
w_i^{k+1} \leftarrow \argmin_{w} \frac{1}{|\data_i|}\sum_{r \in \data_i} \loss(w, r) + \frac{\rho}{2} \TwoNorm{w-z^k}^2 + (w-z^k)^T\mu^{k+1} \label{eqn:primal}
\end{equation}
The dual update in \eqnref{eqn:dual} essentially increases the cost of disagreeing with the consensus value by the constant $\rho$ in the direction of disagreement.
In the synchronization phase of the BSP model, the latest solutions for all $w_i$ and $\mu_i$ are used to recompute the consensus value $z$:
\begin{equation}
z^{k+1} \leftarrow \argmin_{z} \, \lambda  \reg(z) + \frac{z^T p \rho}{2} \left(z - 2 \bar{w}^{k+1} - \bar{\mu}^{k+1}\right) \label{eqn:consensus}
\end{equation}
where $\bar{w}^{k+1}$ and $\bar{\mu}^{k+1}$ are the average values of $w_i^{k+1}$ and $\mu_i^{k+1}$ across all the machines.
For most regularization functions \eqnref{eqn:consensus} can be computed analytically.
For example in the commonly used $L_2$ regularization ($\reg(z) = \TwoNorm{z}^2$) the solution to \eqnref{eqn:consensus} is simply:
\[
z^{k+1} \leftarrow \frac{\rho  p}{\lambda + \rho p} \left(\bar{w}^{k+1} + \frac{1}{\rho} \bar{\mu}^{k+1}\right)
\]
a weighted combination of the primal and dual averages.


\minihead{Intuition} This somewhat complicated-looking algorithm actually has a simple interpretation.
Each solver is given the original objective function with two useful terms attached.
The former, the \textit{consensus term} ($\frac{\rho}{2}\TwoNorm{w_i-z^k}^2$),
penalizes the new local solution ($w_i$) for deviating from the
previous round's average of solutions, $z^k$. We want the
local sub-problem to be able to move (otherwise, no progress would be
possible). However, to keep the solver from moving too far, ADMM
attaches a quadratic penalty (the exact penalty is scaled by a tuning
parameter $\rho$). In effect, the consensus term limits divergence
by acting as a rubber band: local solvers can move from the previous
average, but they will be \emph{increasingly} penalized for doing so.
The latter \textit{Lagrangian term}, $\mu^k (w_i-z^k)$, pushes solvers towards convergence by effectively tilting the solution space based on prior iterations.
The Lagrangian variable, $\mu^k$, is a vector that acts as a price of deviating and helps direct the local solver.

As further intuition, the ADMM consensus term effectively keeps the
local solvers from deviating too far from one another, while the
Lagrangian helps the local solvers move towards a good solution based
on prior progress. In a sense, ADMM acts like the Escrow transaction
method and the demarcation protocol from traditional transaction
processing: individual solvers ``agree'' not to deviate too far (due
to the quadratic penalty).

\minihead{Implementation} Constructing a ASIP variant of the ADMM algorithm (\badmm as illustrated in \myalgref{alg:badmm}) is relatively straightforward: we again break
down the BSP barrier required between primal and dual stages and allow
solvers to proceed in parallel. After solving its $k$th primal stage,
each local solver performs a \push of its current model and issues a
set of \poll requests to receive other solvers's most recent primal
variables and update the consensus and Lagrangian terms before
continuing.

While ADMM has been studied extensively in the theoretical
literature~\cite{boyd-admm}, its empirical benefits in the distributed
setting have not been well-studied. Moreover, \badmm is reminiscent of
proposals for asynchronous
ADMM~\cite{wei-async-admm,iutzeler-async-admm}, none of which have
been empirically evaluated.

%% file: implementation.tex

\section{Embedding ASIP within Spark}
\label{sec:implementation}

Having introduced the ASIP interface and two applications, we turn our
attention to the problem of actually implementing the ASIP abstraction
within a distributed dataflow system. The relatively simple ASIP
interface exposes a wide design space for underlying implementations;
in this section, we discuss several dimensions of this design space as
well as our experiments implementing a ASIP prototype on top of Apache
Spark.


\subsection{Basic {\large\xchg} and Dataflow Integration}

A primary goal of this work is to introduce asynchronous communication
with existing dataflow systems.  Our proposed answer---the ASIP model,
and, in particular, the use of the \xchg operator---poses challenges
to implementation in a synchronous distributed dataflow system such as
Hadoop or Spark.  One could simulate an \xchg-like operation by
discretely time-stepping via individual BSP
rounds~\cite{grace-async,haloop}. However, this is conceptually at
odds with the asynchronous nature of the tasks we study here and
potentially expensive for fine-grained messaging in a distributed
setting.

Instead, the most straightforward integration with an existing
dataflow system is to provide a communication layer between parallel
dataflow tasks via the local network. One generic strategy for doing
so is to simply open sockets between all concurrent ASIP operators and
provide a thin implementation of the \xchg interface, similar to
implementations of Hadoop AllReduce~\cite{vowpal}. This requires
knowledge of the physical ASIP operator instances, but this
knowledge---modulo fault-tolerance and scheduling, discussed
below---is often available from the cluster scheduler.

Our prototype ASIP implementation resides within Apache Spark, a
memory-optimized dataflow engine~\cite{spark}. In our implementation,
we forego the complexity of a socket-based interface and instead
leverage Spark's \textit{existing} actor-based control channel. That
is, Spark maintains a communications network that is used for cluster
maintenance and scheduling, based on Akka's Actor system. To implement
ASIP \xchg, we register a per-ASIP operator actor instance with each
partition within Spark and obtain actor references for all other
parallel ASIP operator instances. The ASIP actor within Spark
facilitates both \push---via one-way messages to other actors---and
\poll---via an in-actor message queue that is appended to upon message
receipt.  In summary, given an input Spark RDD (a table) partitioned
across machines, we add an additional \texttt{mapPartitions} call to
create and register ASIP actors within Spark for each partition and
then in turn expose the \xchg iterator to the ASIP operator logic.

\subsection{Scheduling Parallel ASIP Operators}

In a departure from traditional cluster scheduling,
asynchronous parallel learning tasks must execute concurrently. That
is, to perform asynchronous learning, operators must be able to
exchange information \textit{as they are running}. Thus, we
adopt a gang-scheduled approach~\cite{gang} to ASIP stage execution, and, for
correct execution, all operators should---in the limit
---be able exchange messages. It is unclear how much asynchronous
algorithms would benefit from ASIP in a non-gang-scheduled environment
(e.g., repeatedly execute a smaller set of isolated ASIP tasks). While
we empirically evaluate the effect of minor delays in concurrent
processing (Section~\ref{sec:stragglers}), understanding the
implications of entirely non-concurrent execution is an interesting
area for future work.

\subsection{Message Routing and Delivery}

As we have discussed, there are a number of possible policies for
message routing within the \xchg iterator. One of the simplest
implementations is simple uniform broadcast between parallel ASIP
operator instances: no additional effort is required to configure
topology nor individual message recipients. This is, in fact, the
design we currently pursue in our prototype and is the basic communication
pattern prescribed by \bsgd and \badmm. However, a range of
alternative protocols, including variants of multicast, point-to-point
communication, and aggregation trees are compatible with the ASIP
model. Conceptually, as in Volcano, ASIP's logical information
exchange primitive enables the system to choose the best physical
instantiation of message routing given the hardware
architecture. However, as networking overhead was not a serious
concern in our empirical analysis, we do not further consider this
optimization here.

Similarly, there is an array of options for providing guarantees on
message delivery, from best-effort delivery to exactly-once and
in-order delivery. In our prototype, we pursue---again---a simple and
pragmatic strategy that we find works well in practice: best-effort
delivery. However, for more demanding analytics tasks, we believe
there is interesting work in understanding the implications of these
more complex delivery policies.



\subsection{Managing Data and Control Flow}

ASIP messaging in a distributed dataflow system incurs several
overheads not present in a single-site system. Notably, message
serialization incurs substantially higher costs than
simple cache-line-based and/or shared-memory communication. In
our prototype, we found that---especially in a JVM-based system like
Spark---the CPU overheads due to object serialization and
deserialization could easily rival the cycles spent on actual
optimization if not carefully used.

To fully leverage the strengths of existing dataflow systems,
the \xchg iterator implementation should \textit{not} be used as
a primary means of \textit{data} transfer. Systems like Spark have
carefully optimized their data transfer code paths to account for bulk
data transfer; as a concrete example, Spark foregoes the use of actors
to transfer RDDs but instead uses a second distribution network that
is also aware of the semantics of task execution
``waves''~\cite{orchestra}. Rather,  programmers should use \xchg
operator as a means of exchanging control messages---periodic
synchronization between concurrent ASIP operators. This also preserves
any native scheduler functionality providing data locality for
individual operator placements. Neither of our ASIP algorithms uses
\xchg for actual data transfer between partitions (e.g.,
re-partitioning training data across machines)---rather, the \xchg
operator conveys information about model updates.

In our prototype implementation of \bsgd, we batch (commutative)
gradient updates (as in a MapReduce combiner~\cite{mapreduce}) to
avoid flooding the network, minimizing \xchg traffic. In our
prototype implementation of \badmm, we rate-limit the sending of primal variable
updates. This reduces the potentially adverse impact of these
additional communication channels.



\subsection{Determinism, Fault Tolerance, Stragglers}
\label{sec:determinism}

A common tenet of modern large-scale dataflow engines (including
Spark) is a requirement for deterministic
execution~\cite{spark,mapreduce}. This decision is a departure from
traditional dataflow engines, which forego mid-query fault
tolerance~\cite{exchange}. Nevertheless, at scale, determinism
simplifies fault tolerance (simply execute another copy of the
operator) and straggler mitigation (again, replicate the
operator). Requiring determinism ostensibly also assists in debugging.

In contrast, ASIP is non-deterministic except under stringent
restrictions on \xchg message delivery order and, until now, we have
not discussed fault tolerance. It might appear that ASIP is hopelessly
at odds with the these systems' operational model. However, in the
context of our asynchronous learning tasks, ASIP is actually quite
compatible with these dataflow engines.

In our ASIP implementation, we leverage two key aspects of our learning
tasks. First, neither \bsgd nor \badmm (nor existing, single-node
asynchronous analytics implementations) is serializable, let alone
deterministic. Second, their use of the \xchg operator naturally lends
itself to task failure and restart: the periodic exchange of model
state provides a means of ``catching up'' with the current cluster
state without having to write any additional failure handling
routines. Specifically, a key property of ASIP in the context of these
algorithms is that the information exchanged via the \xchg iterator is
sufficient---on its own---to (approximately) recover the state of
a lost physical operator. If a particular partition fails (or is
\textit{restarted} due to stragglers), the data on that partition can
be re-loaded and the operator re-started; the excellent convergence
guarantees of dual averaging and ADMM ensure that, despite any
temporary deviation from the partition's predecessor state, the
partition's successor task will eventually converge. Thus, these
algorithms are \textit{statistically fault tolerant}. We empirically
evaluate the effect of restarts in our prototype in the next section, and
demonstrate that, under reasonable delays, system restarts do not
destabilize the ASIP-enabled algorithms.

Many cluster compute tasks are unlike the learning tasks we study here
and will not exhibit this statistical fault tolerance
property. Perennial favorite techniques such as asynchronous
checkpointing and snapshot~\cite{naiad,graphlab} can substitute for
statistical fault tolerance if needed, and, as always, full task
restart is always an option. To provide stronger guarantees on message
ordering (thus enabling determinism), we could employ a stronger
ordering protocol such as atomic broadcast~\cite{atomic} or other
global sequencing layer~\cite{calvin}. Given that the above strategies
are sufficient for the complex analytics tasks we consider here, we do
not consider these alternatives further.

In the event of operator duplication due to straggling,
duplicate partitions may unfairly skew the learning tasks towards the
data in the duplicate partition. One solution to this is to only allow
one partition to \push to \xchg at a time. Another is to perform a
more sophisticated equal-share weighting of duplicate partitions'
\push messages. The former solution is potentially less efficient, but
the latter strategy unfortunately requires additional application
semantics. We currently adopt the former, which we also evaluate
in the next section.

%% file: evaluation.tex

\section{Empirical Study of Routines}
\label{sec:evaluation}

To assess the impact of asynchrony within the ASIP model in the context of convex programming, we implemented and evaluated the \bsgd and \badmm algorithms using ASIP on top of Spark.
We compare against the corresponding synchronous algorithms implemented directly in the native Spark dataflow abstraction.
Where possible, we shared the same subroutines and data representations to isolate the gains due to the asynchrony and the ASIP model from variations in code quality and optimizations.

As our goal in this work is to study the costs and benefits of asynchronous convex programming routines in general purpose dataflow systems, we focus our evaluation on our ASIP-based convex programming routines within Apache Spark. Accordingly, we explicitly do not compare against special purpose, non-dataflow computation platforms (e.g., GraphLab). We expect these specialized systems to outperform both Spark and our prototype ASIP implementation for each system's specialized tasks (e.g., graph computation in GraphLab). Our objectives in the section are to:
\begin{evalenumerate}
\item Demonstrate that, for well-behaved inputs, the asynchronous algorithms \bsgd and \badmm implemented using the ASIP model demonstrate relative speedups compared to their corresponding dataflow-based BSP implementations.
\item Expose a trade-off between iteration speed and stability defined by the \bsgd and \badmm algorithms as well as their dependence on specific model parameters.
\item Evaluate the adverse impact sof stragglers and machine failures on the statistical fault tolerance of \bsgd and \badmm.
\end{evalenumerate}

The performance and complexity of each convex programming task is intimately connected to the size, dimension, and signal of the underlying data as well as the properties of the loss and regularization functions. Accordingly, we evaluate each algorithm on four different publicly available datasets using four combinations of loss and regularization functions.
In addition, to understand the effect of skewed data placement, we also evaluate each algorithm on specially crafted synthetic datasets.

To the best of our knowledge, this is the first comprehensive empirical evaluation of distributed synchronous and asynchronous implementations of convex programming algorithms on real-world large-scale tasks using a commodity cluster computing framework.


\subsection{Experimental Setup}

We deployed our prototype a cluster of 16 publicly available EC2 \texttt{m2.4xlarge} worker instances in the Amazon \texttt{us-west-2} (Oregon) region.
Each instance had $68.4$ GB of RAM and an eight-core Intel Xeon E5-2665 running at $2.4$ GHz and was connected to a commodity Gigabit network.
While virtualization can lead to some variability in network and processor performance, we encountered limited variability during our system evaluation.


In our experiments, we primarily report the total objective value \eqnref{eqn:obj} (training loss plus the scaled regularization penalty) as a function of time.
While test error is often a more common metric for machine learning tasks, it conflates modeling with the inferential task (estimating the model parameters) that is the focus of convex programming and the algorithms we study.
Moreover, by selecting a model (loss function, regularization function, and regularization weighting parameter $\lambda$), the remaining inferential task is entirely governed by the objective function defined in \eqnref{eqn:obj}.

To isolate data-loading costs from the algorithm and system comparison (fixed across algorithms), we exclude the time required to load, format, and appropriately partition the raw input data from HDFS.  However, due to optimizations in Spark which exploit data locality and in-memory caching, initialization and data-preprocessing costs were relatively minimal (on the order of 10s of seconds) across all experiment configurations.  This speaks to the advantage of an integrated systems architecture that leverages existing engineering effort in data-centric computation.

Our default (and vendor recommended) Spark configuration launches a separate worker thread for each core on each machine, leading to a total of 128 active worker threads.
As a consequence, we used 128 \xchg iterators, one per core, with all in-phase communication mediated through the ASIP interface.
To ensure fair work balance, we evenly distribute the input records across the 128 worker threads for all datasets.
However, for the synthetic datasets, we adversarially assign records to machines based on the record values, leading to an even work balance, but with substantial statistical imbalance in the values (see \secref{sec:syntheval} for details).

\subsection{Algorithms for Comparison}
\label{sec:algcomparison}

In our analysis, we compare the two asynchronous algorithms, \bsgd and \badmm, with their BSP counterparts---gradient descent (\texttt{GD}) and alternating direction method of multipliers (\texttt{ADMM})---as well as a \naive averaging algorithm (\texttt{AVG}).  \texttt{AVG}, is perhaps the simplest distributed inference algorithms and divides the problem among each of the machines and then applies a local fast SGD solver to solve the problems independently, averaging the final solution.  The \texttt{AVG} algorithm, advocated by \cite{bismarck, smola-averaging}, provides an intuitive coordination-avoiding baseline.

The closest BSP analog to \bsgd is (mini)batch gradient descent.
This algorithm iteratively computes the gradient (of a sample) of the data on each machine, aggregates the gradient across the cluster, and takes a gradient step.
The baseline \texttt{GD} algorithm is taken directly from the open-source machine learning library (MLlib) built into Spark~\cite{mli} and relies on several internal optimizations including parallel reduction trees for more efficient aggregation.
We used the default minibatch size suggested by Spark, which is the full dataset size.
This is because the Spark sampling routines require a full scan of the data, which is close to the cost of the gradient calculation.

Our implementation of the ADMM algorithm directly follows the description in \secref{sec:admm}.
As a local solver for both ADMM and \badmm, we used mini-batch SGD (Pegasos SVM~\cite{pegasos}) with an $\eta/\sqrt{t}$ decreasing step size.
This is the same step size configuration used for \bsgd.
The mini-batch SGD algorithm computes the gradient of the sub-problem with respect to fixed-sized sample of local records and then applies a gradient step.
In contrast to vanilla SGD (batch size 1), the mini-batch improves the gradient estimator by averaging over several records.
We found that a batch size of 10 to 100 records generally performed well.  In our experiments we used a batch size of 10 records.

The machine learning literature \cite{boyd-admm} suggests the use of several more sophisticated sub-problem solvers for ADMM (i.e., instead of SGD).  We experimented with gradient descent using backtracking line-search as well as the limited memory variant of the Broyden-Fletcher-Goldfarb-Shanno (l-BFGS) algorithm~\cite{lbfgs-ref}.  While these techniques slightly improved the local objective value and stability they substantially increased wall-clock runtime, generally leading to poor performance as a function of time.

One of the surprising challenges in our experimental deployment was the algorithms' sensitivity to the choice of constants (i.e., \textit{hyperparameters}).  Both the ADMM quadratic penalty $\rho$ and the SGD step-size $\eta$ had a measurable impact on the rate of convergence and stability.  We chose a single, reasonably robust setting for these parameters that we applied uniformly across the datasets and objective formulations (see Appendix~\ref{sec:appendix-exp}, Table~\ref{table:params}).  Automatically tuning these parameter for different workloads and perhaps even cluster configurations would likely lead to improved performance and stability at the cost of complexity.

\subsection{Comparison on Real-World Datasets}

\begin{figure*}[ht]
\vspace{-1em}
\textbf{\hspace{.6in}SVM+L2 \hspace{1.22in} SVM+L1 \hspace{1.22in} LR+L2 \hspace{1.25in} LR+L1\hspace{.75in}}\\
\includegraphics[width=1.7in]{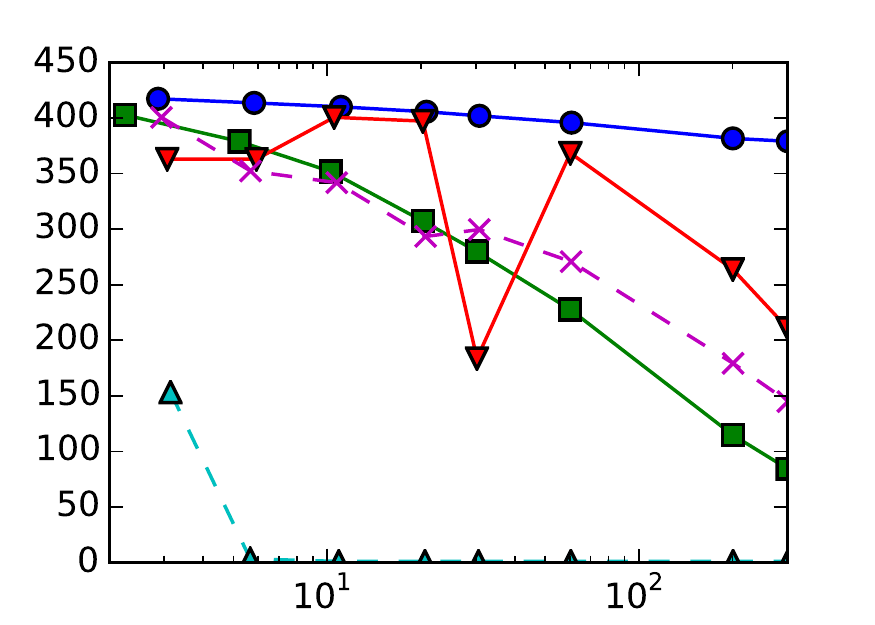}
\includegraphics[width=1.7in]{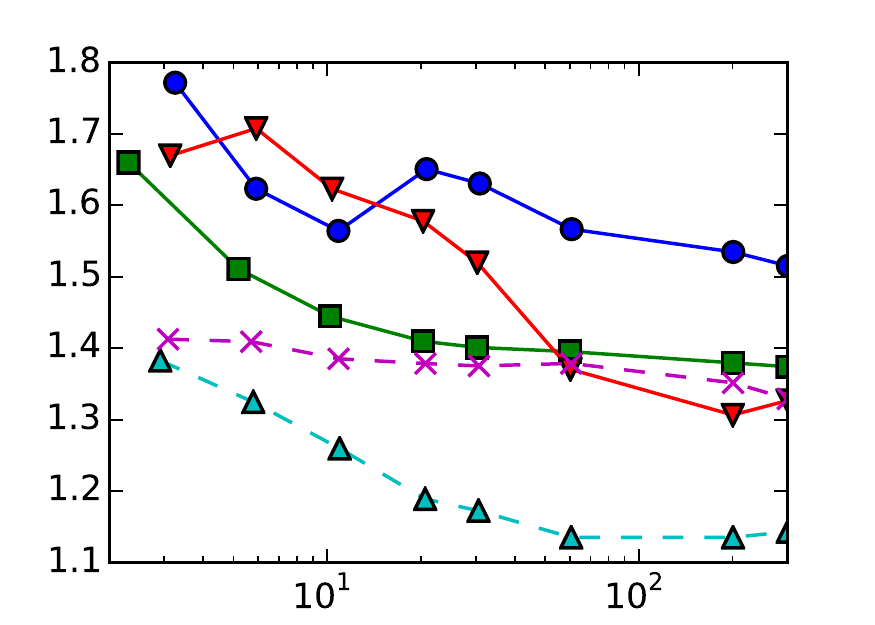}
\includegraphics[width=1.7in]{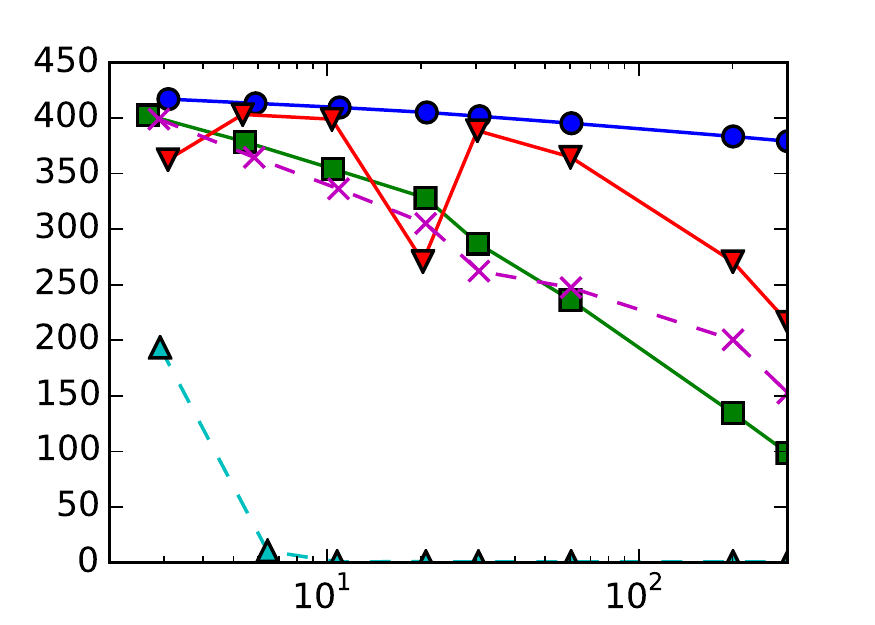}
\includegraphics[width=1.7in]{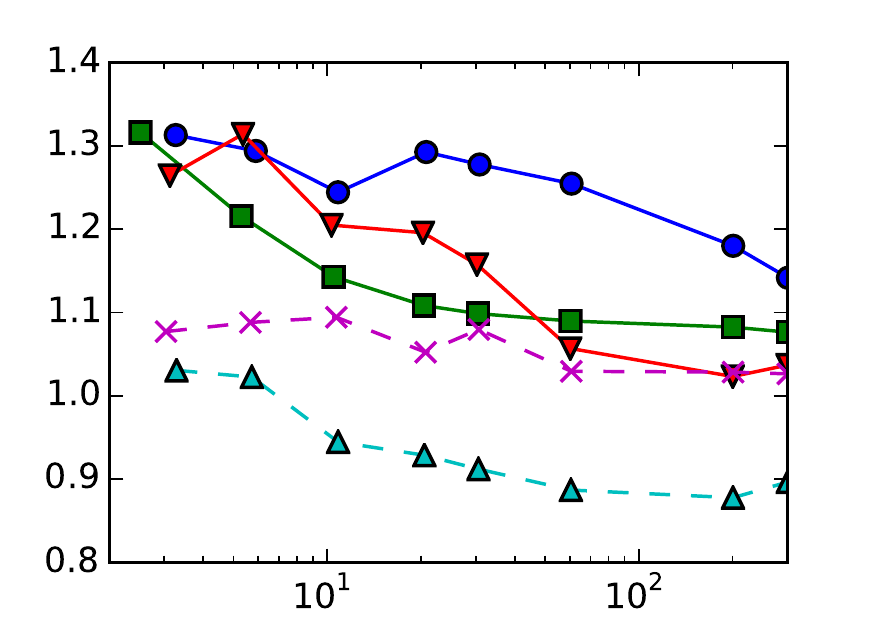}\vspace{-.75em}
\begin{center}\textbf{flights}: 7,009,728 records, dimension 8171,
  2.4GB RDD (sparse), 688MB raw\end{center}
\includegraphics[width=1.7in]{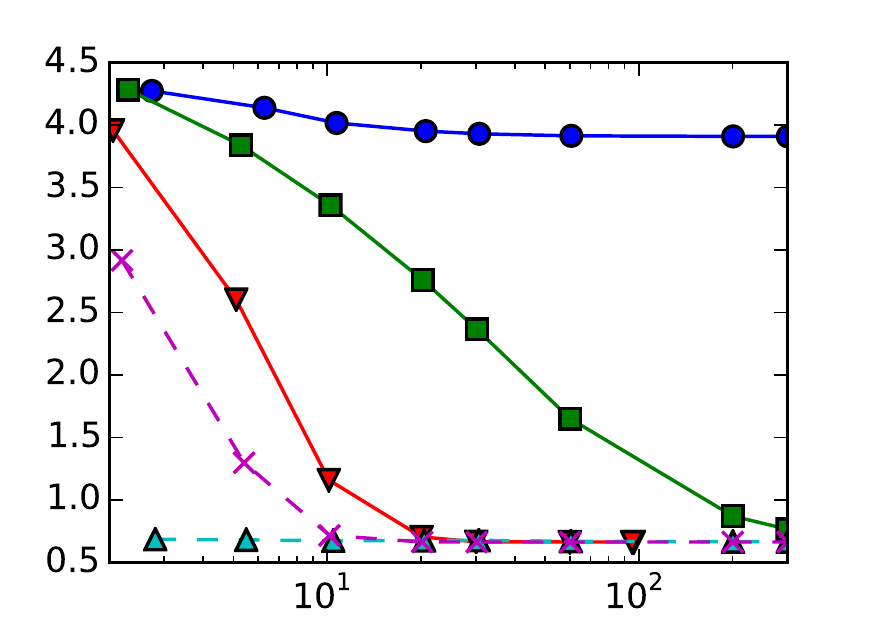}
\includegraphics[width=1.7in]{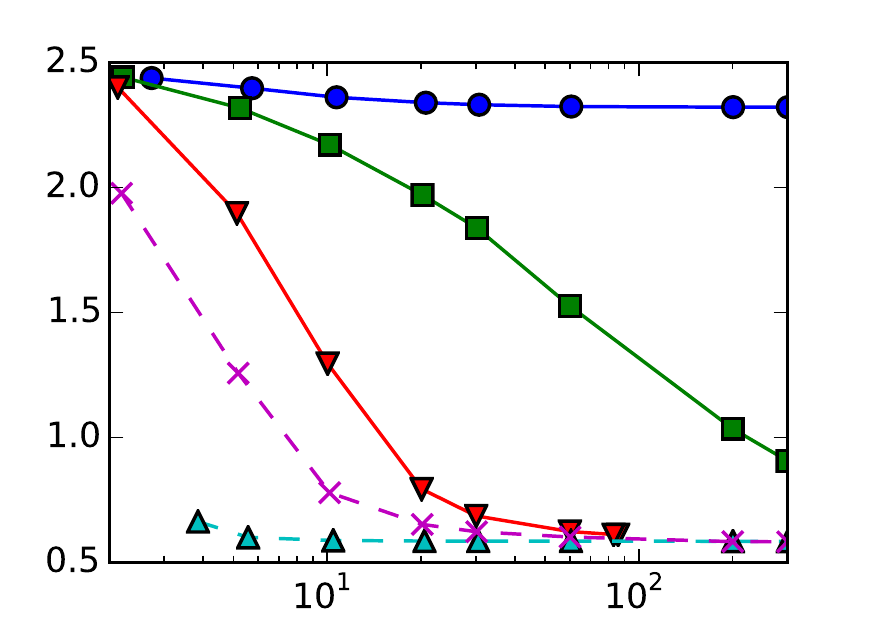}
\includegraphics[width=1.7in]{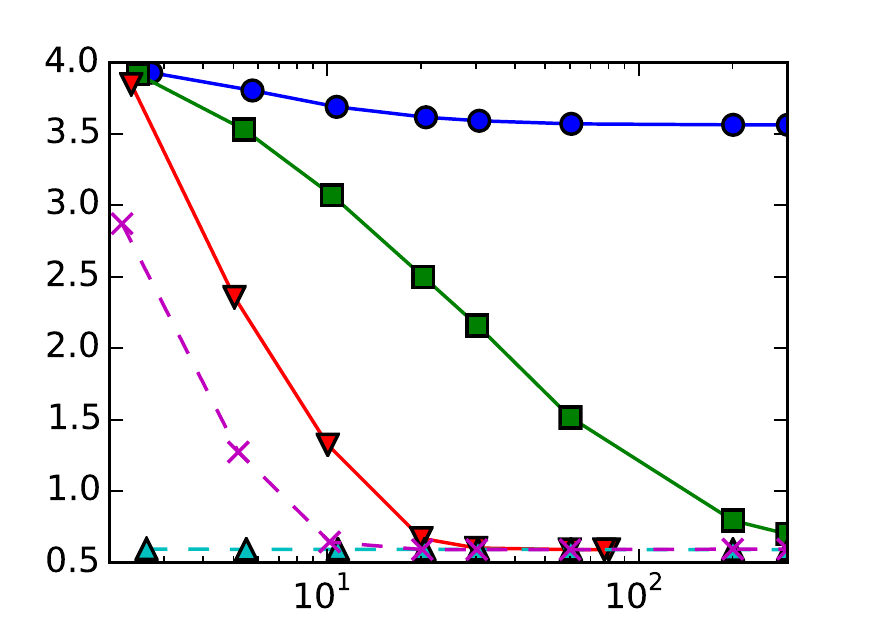}
\includegraphics[width=1.7in]{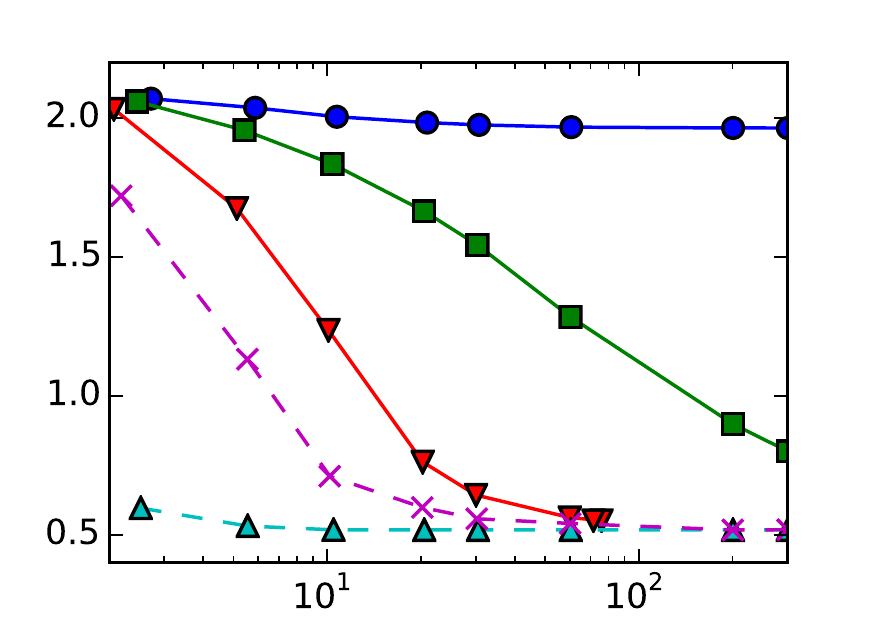}\vspace{-.75em}
\begin{center}\textbf{forest}: 581,012 records, dimension 54,
  350.2MB RDD, 507.5MB raw\end{center}
\includegraphics[width=1.7in]{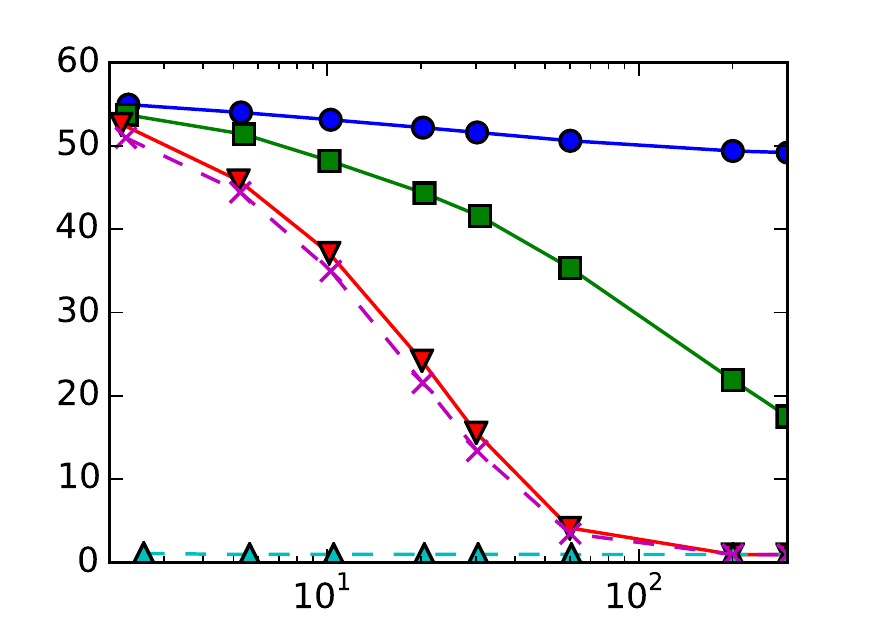}
\includegraphics[width=1.7in]{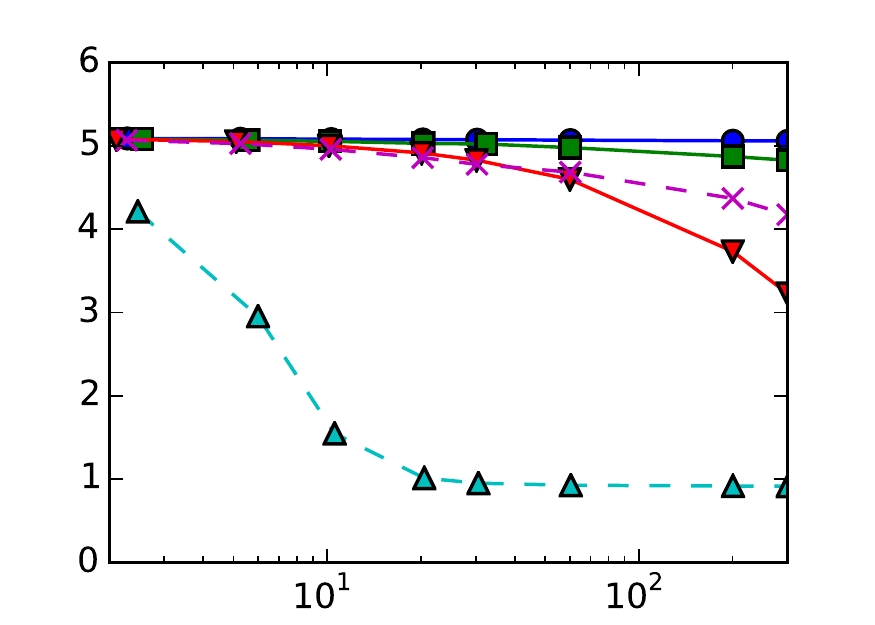}
\includegraphics[width=1.7in]{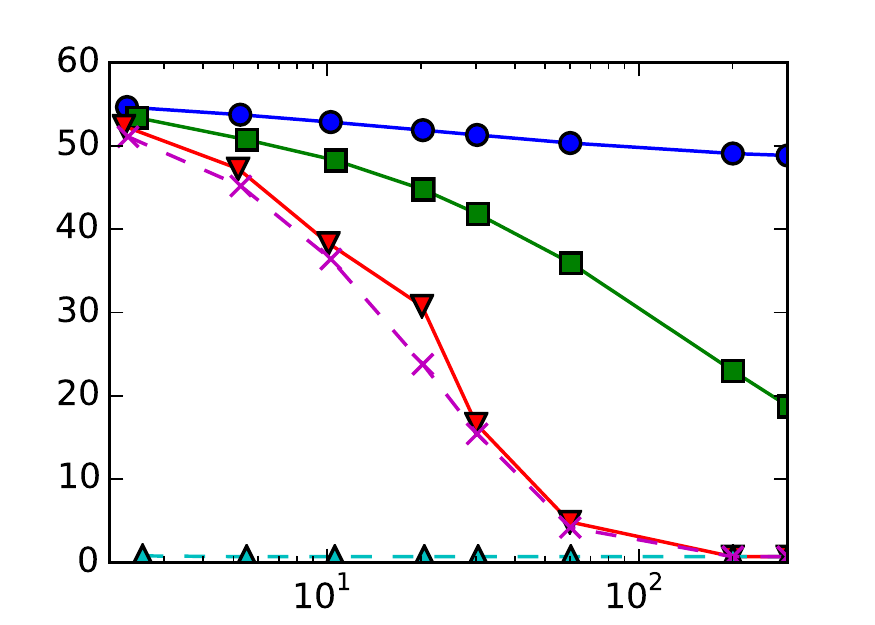}
\includegraphics[width=1.7in]{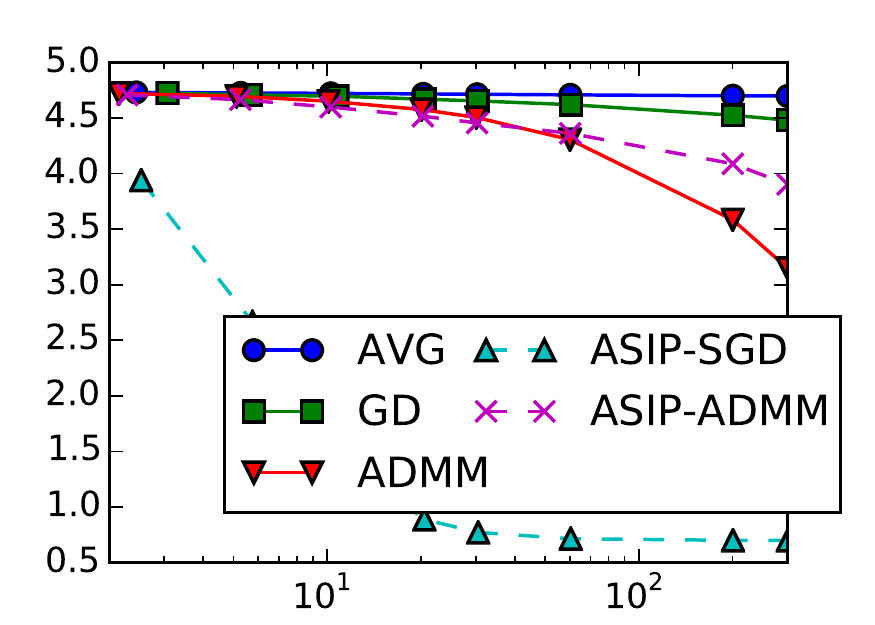}\vspace{-.75em}
\begin{center}\textbf{wikipedia}: 6,745,869 records, feature hashing
  to dimension 1000, 51.5GB RDD, 49.0GB raw\end{center}
\includegraphics[width=1.7in]{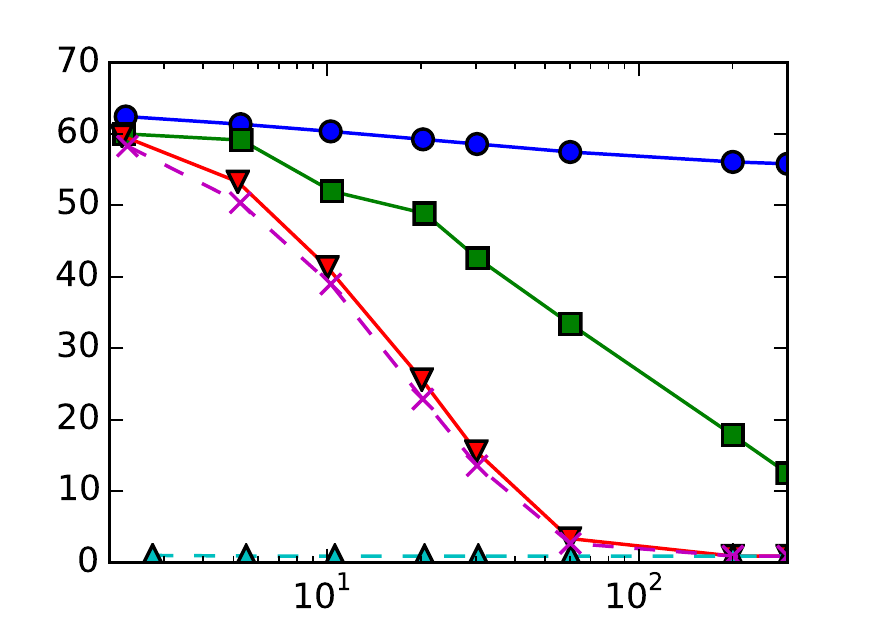}
\includegraphics[width=1.7in]{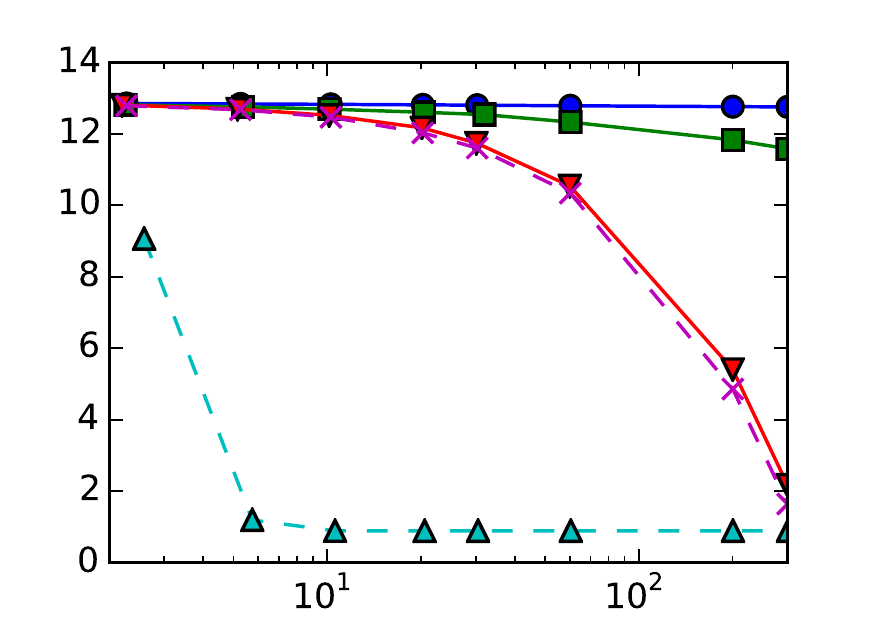}
\includegraphics[width=1.7in]{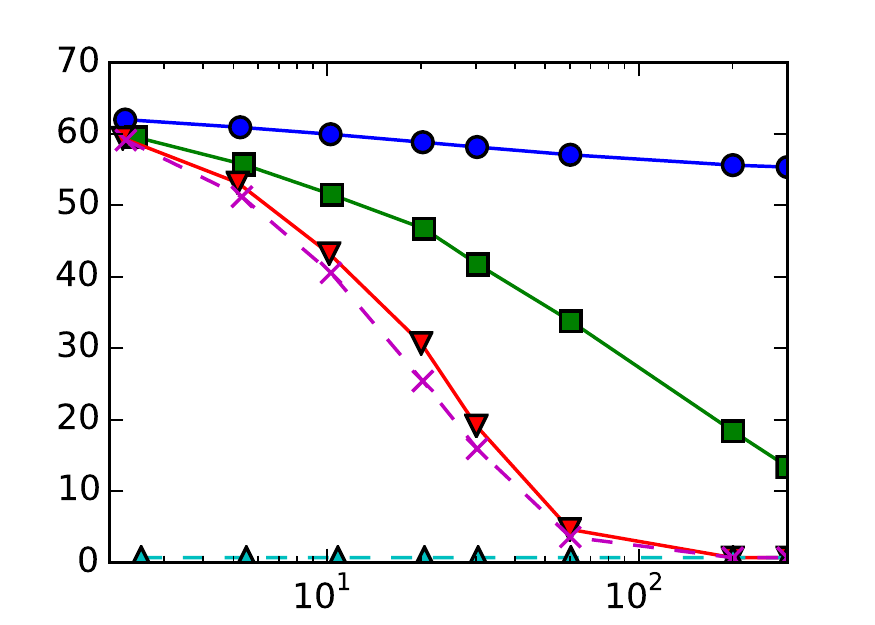}
\includegraphics[width=1.7in]{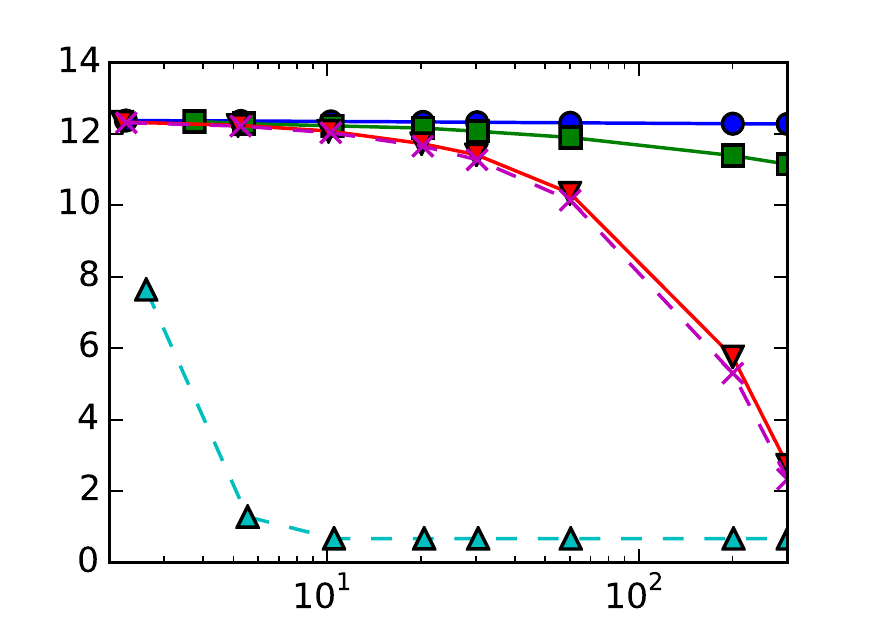}\vspace{-.75em}
\begin{center}\textbf{DBLP}: 2,704,455 records, feature hashing to dimension 1000,
  20.7GB RDD, 1.5GB raw\end{center}\vspace{-1em}
\caption{\footnotesize Evaluation on Real-World Datasets. We plot the objective value \eqnref{eqn:obj} versus runtime (in seconds) for four real-world datasets using four variations on the objective function. In general we find that \bsgd out-performs all other algorithms though is less stable. Alternatively, \badmm is more stable and generally out-performs its BSP counterpart. }\vspace{-1em}
\label{fig:realcomparison}
\end{figure*}

To compare the performance of each algorithms, we evaluated them on a range of datasets and training tasks (\figref{fig:realcomparison}).
We consider four real-world publicly available datasets ranging in both size and dimensionality.
\begin{myitemize}
\item The \texttt{flights}\cite{flights} dataset consists of over 7M airline flight records in 8K dimensions describing one-hot encoded characteristics of each flight and whether it was late.
As a consequence of the one-hot encoding the \texttt{flights} is highly sparse and requires only 2.4GB to store in the memory of the cluster.
\item The \texttt{forest}\cite{forest} dataset (evaluated in~\cite{bismarck}), which relates characteristics of regions of a forest to whether they burn in a forest fire, is the smallest dataset we used, at only half a million records and 54 dimensions.
\item The largest dataset was \texttt{wikipedia} at 6.7M records in 1000 dimensions (dense).  The \texttt{wikipedia} dataset was constructed by applying feature hashing~\cite{Weinberger09} to encode the bag-of-words representation of each article in a dense 1000 dimensional space and then predict the presence of the word \emph{database} in each article.
\item Finally, the \texttt{DBLP} \cite{dblp} consists of 2.7M records which, like \texttt{wikipedia}, were constructed by hashing the title of each article in DBLP using bag-of-words representation to 1000 dimensional feature space. We predicted, based on the title, whether an article was written before or after 2007.
\end{myitemize}

In addition to the four real-world datasets, we also evaluated four different learning tasks, achieved by modifying the solver objective functions.  We constructed each function by combining one of two common classification loss functions:
\begin{align}
\text{Hinge Loss: } & \loss(x, r = (f, y)) := \max(0, 1 - y  x^T f) \\
\text{Logistic Loss: }  & \loss(x, r = (f,y)) := \log\left( 1+ \exp(-y x^T f)\right)
\end{align}
corresponding to support vector machines and logistic regression, respectively, with two common regularization functions:
\begin{align}
L_1: \quad &  \reg(x) := \OneNorm{x} \\
L_2: \quad & \reg(x) := \frac{1}{2} \TwoNorm{x}^2
\end{align}
corresponding to the Lasso and Tikhonov regularization~\cite{boyd-book}.  In terms of the complexity of the optimization objective, we expect the hinge loss and $L_1$ regularization function to be the most challenging as they are non-smooth and encourage sparse solutions requiring greater coordination.  While it is not reasonable to compare the objective value across objective functions, we can compare the relative performance of each algorithm within each learning task.

Across all the combination of objective function and datasets we observe (\figref{fig:realcomparison}) a few common trends.  In general, \naive averaging performs poorly, while the \bsgd generally substantially outperforms the other algorithms.  In general, the asynchronous variants \bsgd and \badmm out performed their synchronous counterparts, with \bsgd often outperforming \texttt{GD} by more than an order of magnitude.
In particular, if we consider the average ratio of the objective at 10 seconds into the computation across all experiments we find that \bsgd is 74 times lower than that of \texttt{GD}.
While the objective of \badmm is only 1.2 times lower than that \texttt{ADMM} we show in \figref{fig:stragglers} that in the presence of a single straggler \badmm can yield more than an order of magnitude reduction in objective value.

The choice of objective value had a noticeable effect on the overall convergence.
In general, we saw greater variability in the final objective value for $L_1$ regularization, especially when combined with the SVM.
Because the $L_1$ objective seeks a \emph{sparse} solution, it requires greater coordination across processors for convergence, and, therefore, we would expect algorithms that exploit ASIP communication to converge faster.
Indeed, when the $L_1$ objective was used on the two dense high-dimensional text datasets, we found that the ADMM based techniques generally performed poorly while \bsgd was able to quickly attain a much lower objective value.


\subsection{Point Cloud: The Effect of Data Skew}
\label{sec:syntheval}

\begin{figure*}[ht]
\vspace{-1em}
\subfigure[Point Cloud Dataset]{
\includegraphics[width=.23\linewidth]{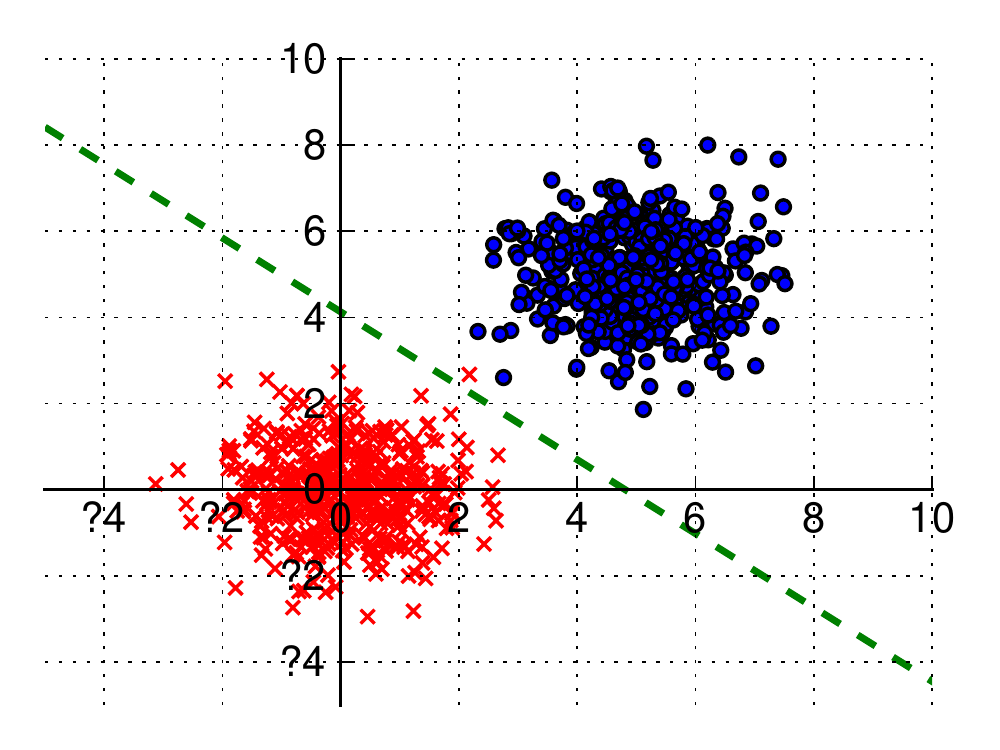}
}
\subfigure[ADMM Iteration 2]{
  \includegraphics[width=.23\linewidth]{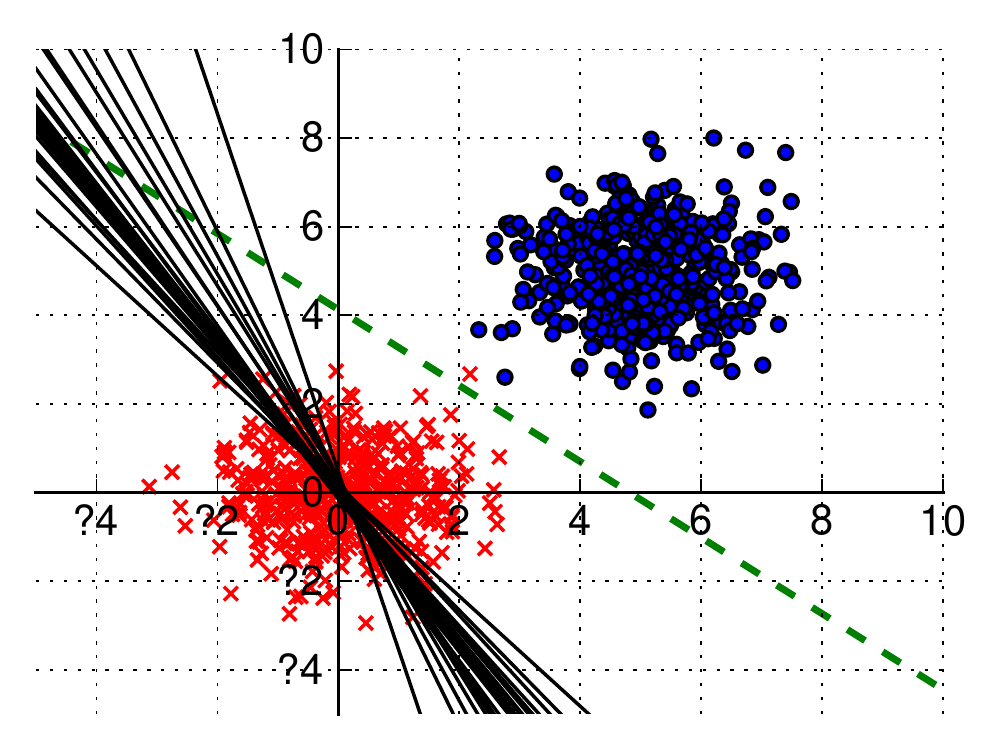}
}
\subfigure[ADMM Iteration 10]{
  \includegraphics[width=.23\linewidth]{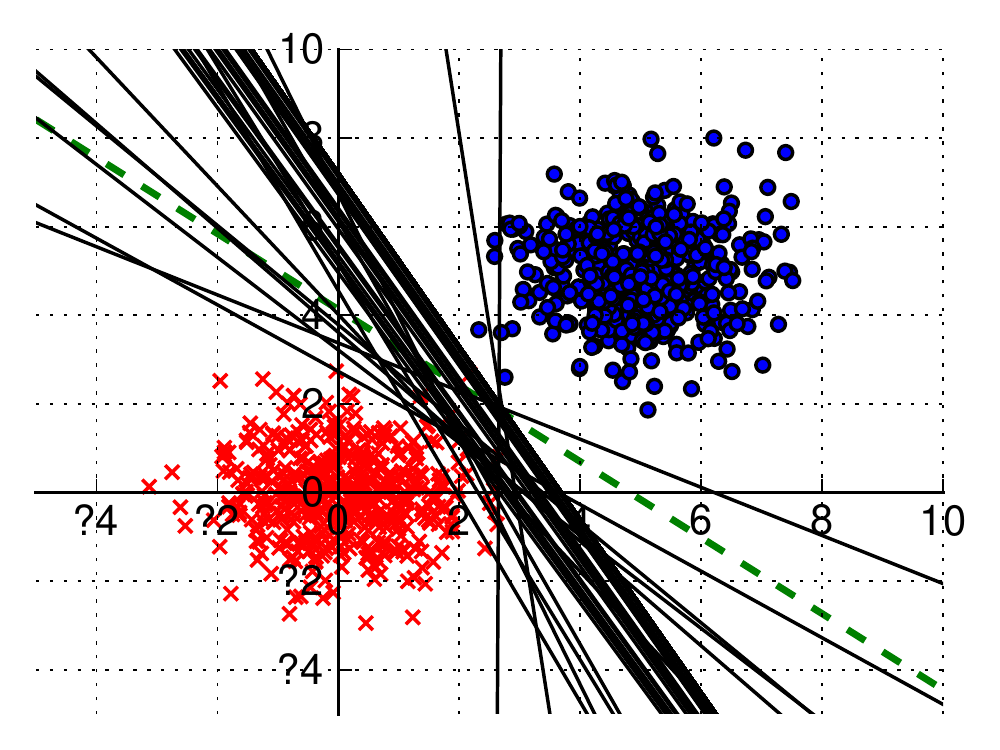}
}
\subfigure[ADMM Iteration 20]{
  \includegraphics[width=.23\linewidth]{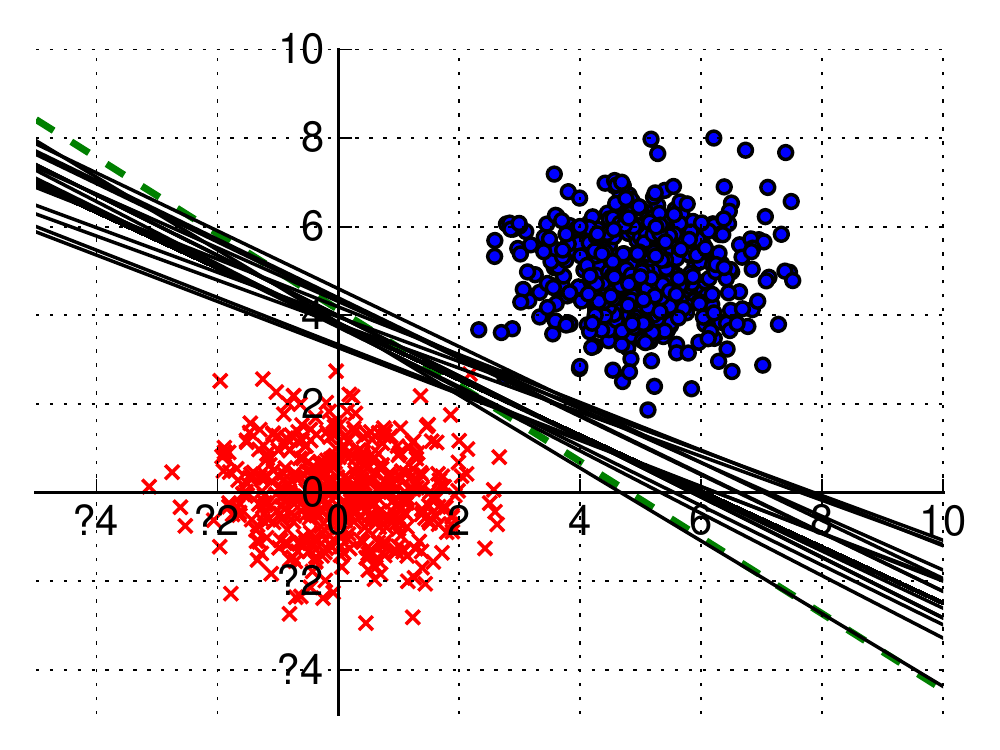}
}\vspace{-1em}
\caption{\footnotesize Point Cloud Data and Hyperplanes. Here we illustrate the point cloud data set with two classes centered at $(0,0)$ and $(5,5)$.  The later sequence of plots illustrate the execution of the ADMM algorithm. Each of the black lines corresponds to 1 of the 128 hyperplanes computed on each of the \xchg iterators.
}
\label{fig:cloudsim}
\end{figure*}

\begin{figure*}[ht]
\textbf{\hspace{.45in} Skewed SVM+L2 \hspace{.75in} Skewed LR+L1 \hspace{.75in} Uniform SVM+L2 \hspace{.75in} Uniform LR+L1\hspace{.75in}}\\
\includegraphics[width=1.7in]{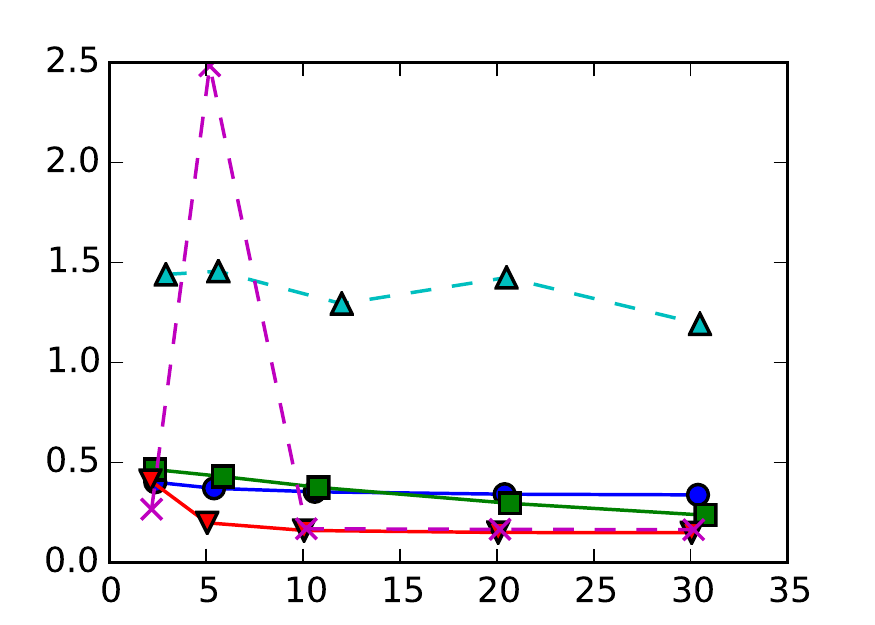}
\includegraphics[width=1.7in]{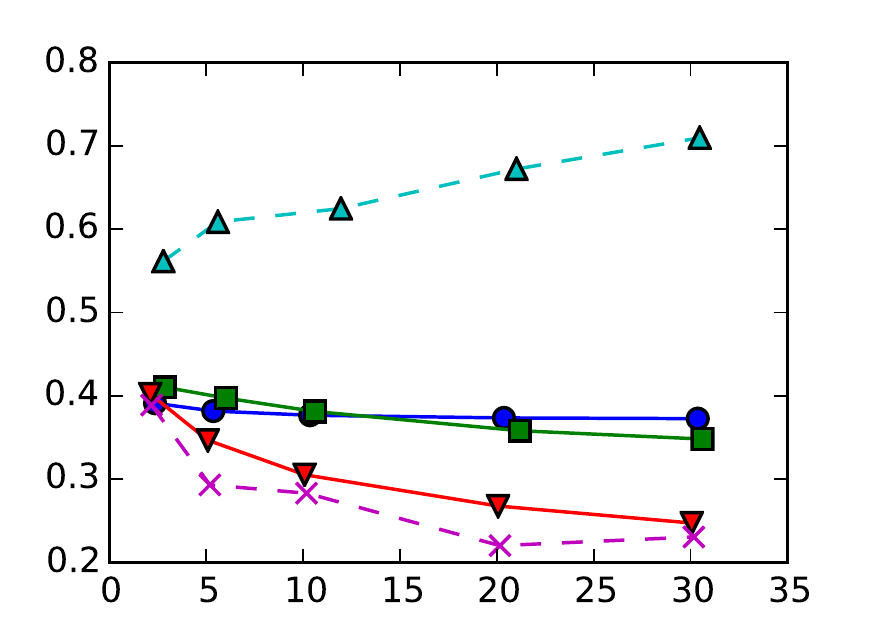}
\includegraphics[width=1.7in]{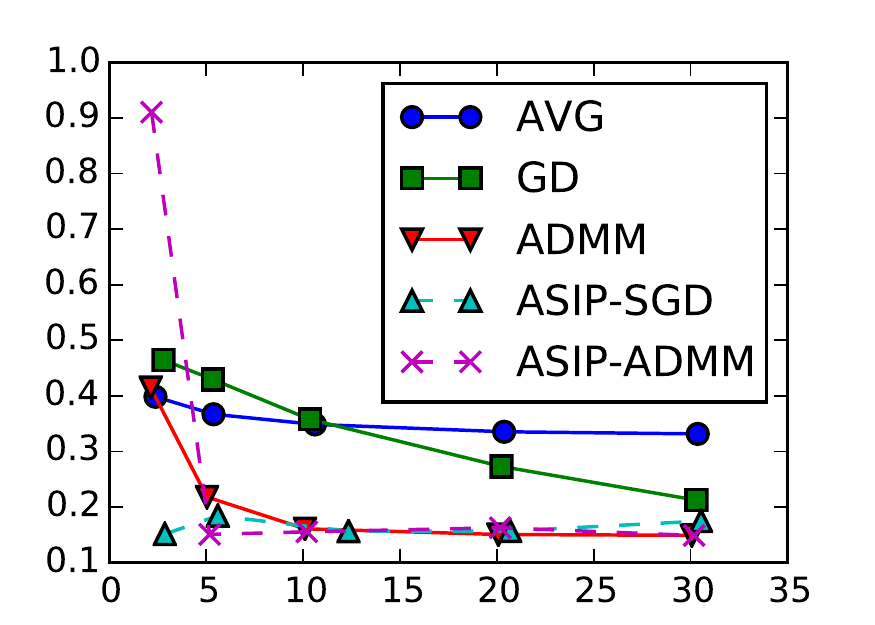}
\includegraphics[width=1.7in]{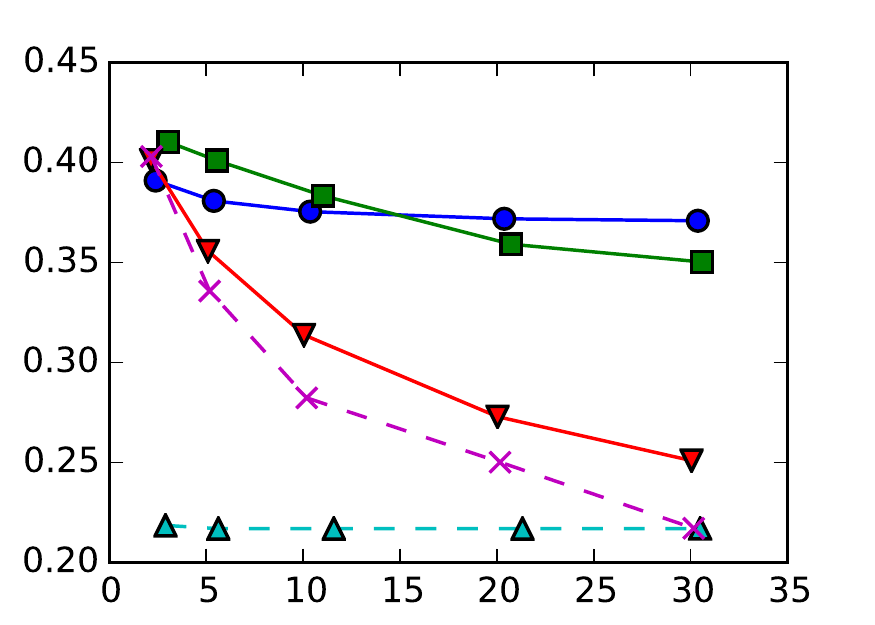}\vspace{-.75em}
\caption{\footnotesize Comparison of Skewed Synthetic Data.  We plot the objective value \eqnref{eqn:obj} versus runtime (in seconds) comparing skewed data placement with uniform data placement. In general, we find that highly skewed data-placement causes the \bsgd algorithm to diverge.}
\label{fig:synthcomparison}
\end{figure*}

To evaluate the effect of a skewed data placement, we introduced a synthetic dataset with a well understood optimal solution.
The synthetic dataset (see \figref{fig:cloudsim}) consists of two ``point clouds,'' with Gaussian distributions centered at (0,0) and (5,5).
Because this dataset is not separable with a hyperplane passing through the origin, we also introduce an additional bias dimension (achieved by setting the third dimension of each point cloud to one).
To evaluate skew, we assigned each machine data points from only one of the two clouds leading to extreme \emph{label bias}.
As a consequence, using only the data available to that machine, it is \textit{impossible} to recover the optimal solution---communication is required..
Thus, the particular coordination protocol between machines is essential to computing the correct answer.
While in general such extreme skew is unlikely, we believe that for sparser datasets and task specific data partitioning, it is possible that similar skew could be observed in real-world workloads (e.g., users partitioned by zip code might introduce a strong bias on a feature such as annual income; this is the ``CA-TX'' problem of~\cite{bismarck}).

In \figref{fig:synthcomparison}, we plot the objective value of each of the algorithms as a function of time for the skewed data placement and the uniform random placement. We show only the \texttt{SVM+L2} and \texttt{LR+L1} objectives since these are the most common objectives (although a similar behavior was observed for the other objectives).  In contrast to the results on the real-world datasets, we find that \bsgd generally performs worse on skewed data.  This discrepancy is due in part to the effect of delays in communication and imbalance in communication across machines playing a key role in the final solution.  Conversely, the more robust behavior achieved by \badmm leads to relatively stable convergence.  Interestingly, the spike in objective at 5 seconds in the execution of \badmm on the skewed dataset is not noise but actually an artifact of a temporarily suboptimal solution as a fraction of machines must flip their solutions to achieve consensus.




\subsection{Evaluating Straggler Overheads}
\label{sec:stragglers}

Stragglers are a surprisingly common phenomena, even on isolated workloads~\cite{mr-fnt,spark,mapreduce}.
For example, JVM cluster compute frameworks such as Spark often experience long garbage collection pauses.
Furthermore the iterative nature of convex programming algorithm only exacerbates the affect of stragglers.

\begin{figure*}[ht]
\subfigure[forest]{
  \includegraphics[width=.23\linewidth]{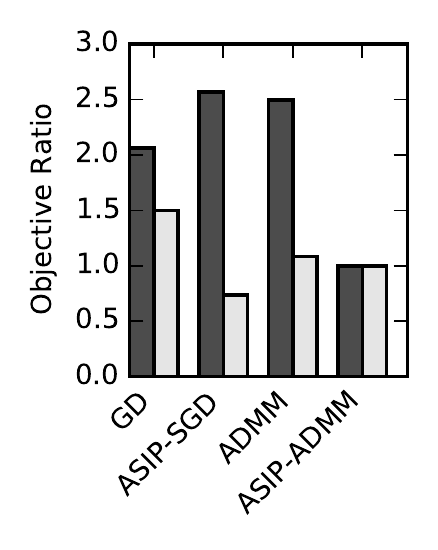}
}
\subfigure[flights]{
  \includegraphics[width=.23\linewidth]{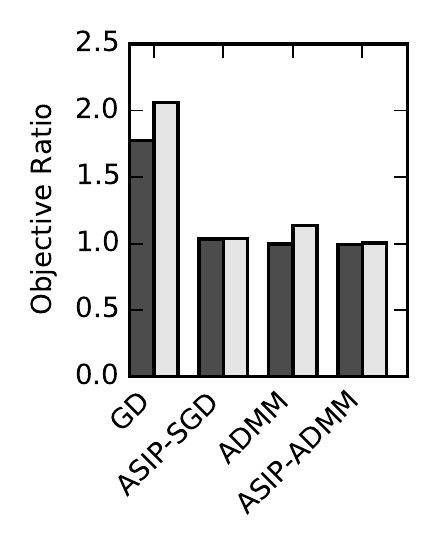}
}
\subfigure[dblp]{
  \includegraphics[width=.23\linewidth]{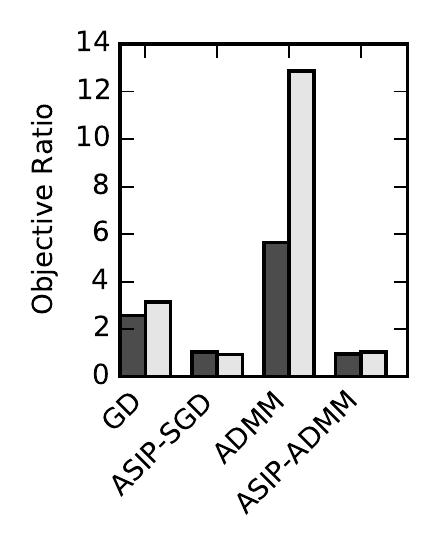}
}
\subfigure[wikipedia]{
  \includegraphics[width=.23\linewidth]{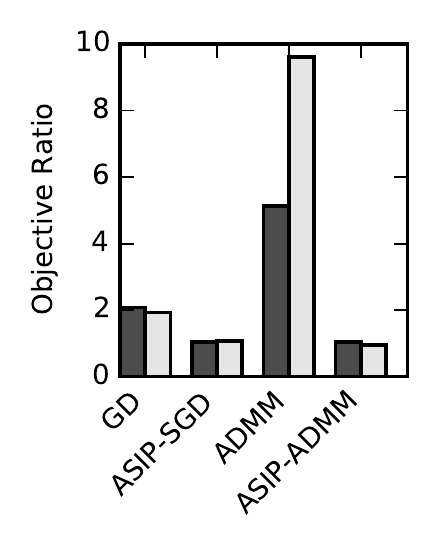}
}\vspace{-1em}
\caption{\footnotesize Straggler Experiments.  Here we assess the affect of stragglers by introducing a period pause in one of the worker cores to simulate a 1 second GC every two seconds. We plot the ratio of the objective (\texttt{SVM+L2}) with the period pause to the objective without pauses (lower is better).  The dark shaded bar is the objective ratio at 5 seconds and the light bar is at 10 seconds.
As expected the BSP algorithms are sensitive to stragglers and in some cases resulting in order of magnitude reductions in overall performance.
}
\label{fig:stragglers}
\end{figure*}

One of the primary advantages of the ASIP model is that it is inherently robust to stragglers.
By eliminating points of blocking coordination, the ASIP model mitigates the affect of slower processors. However, these benefits must be weighted against the possible statistical imbalance that stragglers might introduce.
To assess the effect of stragglers on the overall objective in a controlled setting, we introduced a synthetic one second pause every two seconds in just one of the cores in our cluster.
In \figref{fig:stragglers} we plot the ratio of the objective with pauses to that without pauses at 5 seconds (dark bar) and 10 seconds (light bar).
In general, we notice that the BSP algorithms tend to perform poorly and in some cases are an order of magnitude worse than the ASIP algorithms; in the ASIP algorithms, the straggling processes do not cause the non-straggling processes to stop processing data.
Interestingly, in the smaller \texttt{forest} dataset, where the optimal solution is obtained quickly, the introduction of a straggler can temporarily destabilize the solution, but, by 10 seconds, the optimal value is recovered.

\subsection{Evaluating Statistical Fault Tolerance}
\label{sec:fault-tolerance}

As we discussed in Section~\ref{sec:determinism}, one of the key features of synchronous dataflow systems like Spark are that they typically assume tasks are deterministic, enabling logical logging based fault-tolerance; this observation forms the foundation of the Spark fault tolerance model.
Conversely, as we have discussed, ASIP is non-deterministic, but because ASIP iterators frequently share state, the \emph{important} state of the system in our algorithms is already \emph{replicated}.
Furthermore, the convex programming algorithms are relatively robust to perturbations in the state of the model enabling fast recovery from the replicated, inconsistent, state on other machines (i.e., are statistically fault tolerant).

\begin{figure*}[ht]
\subfigure[forest]{
  \includegraphics[width=.23\linewidth]{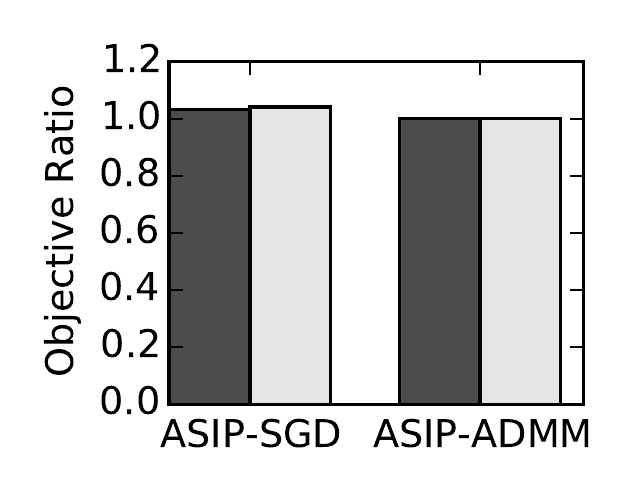}
}
\subfigure[flights]{
  \includegraphics[width=.23\linewidth]{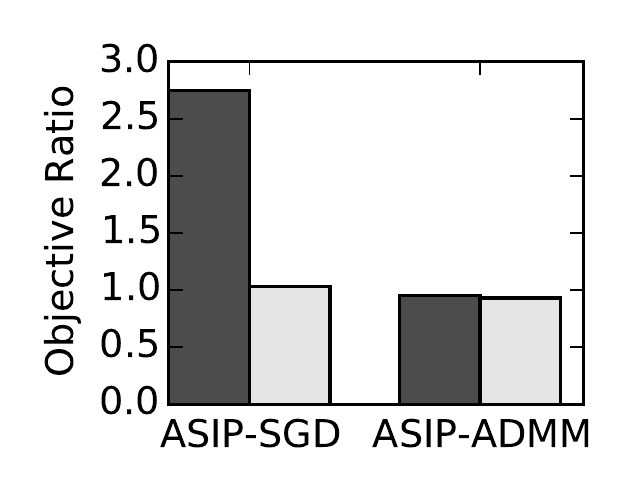}
}
\subfigure[dblp]{
  \includegraphics[width=.23\linewidth]{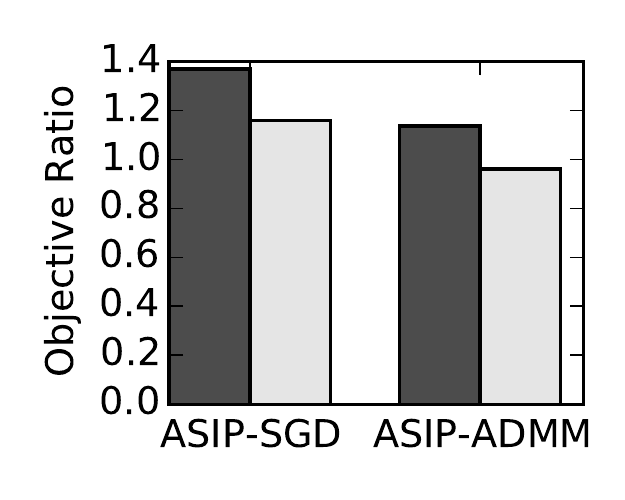}
}
\subfigure[wikipedia]{
  \includegraphics[width=.23\linewidth]{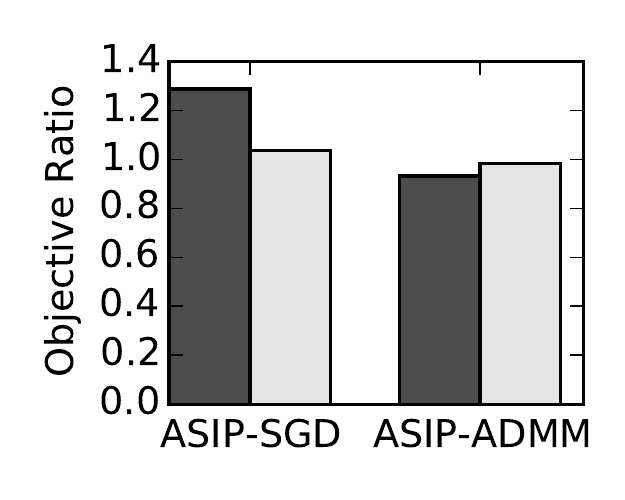}
}\vspace{-1em}
\caption{\footnotesize Fault Tolerance Experiment. We plot the ratio of the objective (\texttt{SVM+L2}) after a machine reset 3 seconds into the computation and the objective without reset (lower is better).  The dark shaded bar is the objective ratio at 5 seconds and the light bar is at 10 seconds.  In general we find that the asynchronous algorithms are relatively robust to machine failure.}
\label{fig:faulttolerance}
\end{figure*}

To determine the degree of statistical fault tolerance under machine failure, we simulate the failure and relaunch of a machine (8 iterators) in our cluster.
Using the \texttt{forest} dataset, we inject a machine failure after the initial convergence of 3 seconds and plot (\figref{fig:faulttolerance}) the relative objective at 5 and 10 seconds (lower is better).
The smaller timescales reflect the relatively fast convergence of the asynchronous techniques.
Nonetheless, in \figref{fig:faulttolerance-10seconds} (located in the appendix), we consider the effect of introducing error after model convergence (10 seconds) and observe even less of an impact on the objective value at 15 and 30 seconds.
While in practice it may take longer to detect a failure and transition computation to an alternative node, we wanted to focus on the impact on the objective and not the issues related to general cluster management.
In general, we find that the introduction of a node failure has minimal impact in the overall objective, thus demonstrating statistical fault tolerance in this scenario.

%% file: relatedwork.tex

\section{Related Work}
\label{sec:relatedwork}


\minihead{Dataflow systems} Recent years have seen a resurgence of
interest in distributed dataflow
systems~\cite{spark,stratosphere,mapreduce,tupleware,dryad}, both in
research and in industry. These systems continue a long tradition of
expressing large-scale data-parallel computation via dataflow dating
to the earliest relational database systems~\cite{mr-fnt}. Many of
these modern systems, like Spark and MapReduce expose a synchronous
programming model similar to Valiant's BSP abstraction~\cite{valiant-bsp}.

In this work, we examine the problem of allowing fine-grained data
transfer for the purposes of asynchronous analytics within an
otherwise synchronous dataflow system. This problem has several close
relatives in the literature from which we draw inspiration. Volcano's
Exchange operator allows transparent parallelization and distribution
of operators~\cite{exchange,graefe-xchg} (cf. Bubba's non-transparent bracket
model~\cite{bubba}). Our use of \xchg is inspired by Volcano's
exchange but is specifically designed as a specialized iterator for
fine-grained asynchronous model sharing rather than general-purpose,
coarse-grained data transfer. \xchg's
non-blocking receive is similar to Fjords, which enable non-blocking
computation on data streams, provide multi-input iterators, and can
also implement Exchange semantics~\cite{fjord}. Shanmugasundaram et
al. also study the problem of producing partial results in the context
of online aggregation~\cite{ola-partial}.

Cyclic dataflow has been studied in several contexts, including
declarative networking~\cite{dn-implementing}, network
monitoring~\cite{dn-eurosys}, and online query
optimization~\cite{eddies}. Chandramouli et al. propose the
Flying-Fixed-Point operator to track forward progress for cyclic
dataflow~\cite{ffp-dataflow}. Naiad's Timely Dataflow uses an
elaborate, system-wide timestamping mechanism to track progress and to
provide ``consistent,'' partially-ordered outputs for cyclic dataflow
(i.e., fixed-point computation)~\cite{naiad}. In contrast, we target
entirely asynchronous data execution between a set of gang-scheduled
iterators and therefore do not make use of these
techniques. Nevertheless, as we have discussed, efficiently extending
these asynchronous analytics to a non-gang-scheduled environment
appears challenging.

\minihead{In-database analytics} The last several years have also seen
considerable interest in incorporating advanced statistical analytics
into data processing engines at several scales. Within traditional
RDBMS systems, these efforts have centered around bringing advanced
features such as various clustering~\cite{ordonez-clustering} and
classification~\cite{oracle-svm,ordonez-bayesian,gupta-crf,wang-crf,chen-crf}
techniques, Monte Carlo sampling~\cite{mcdb,mcmc-vldb}, and graphical
models and related inference techniques~\cite{sen-graphical,
  wang-graphical,zhang-gibbs} to the database. A related set of
efforts has sought to provide \textit{unified} in-RDBMS processing
capability---that is, common infrastructure for supporting these
tasks~\cite{brown-predictive}, including MADLib~\cite{madlib} and
Bismarck~\cite{bismarck}, which provide high-level user interfaces and
enable easy addition of new algorithms. In parallel, we have seen a
rise of in interest in distributed cluster computation frameworks,
with both custom implementations of particular advanced analytics
algorithms~\cite{panda-trees,scalingml-book,mr-tutorial,mr-set-similarity,sparkler}
as well as general-purpose analytics packages such as
SystemML~\cite{systemml}, MLBase~\cite{mlbase}, MLI~\cite{mli}, Vowpal
Wabbit~\cite{vowpal}, GraphLab~\cite{graphlab}, and
Cumulon~\cite{cumulon} to facilitate them. Our goal in this work is to
efficiently support \textit{asynchronous} complex analytics in a
generic, distributed dataflow environment.


\minihead{Convex programming and asynchronous algorithms} The problem
of distributed convex optimization has a long history in the
literature~\cite{boyd-book}, predating the development of relational
algebra by several decades. There are a variety of techniques from
this literature that have received considerable theoretical attention
and some experimental evaluation; in this paper, we discussed and
evaluated several, including distributed
averaging~\cite{smola-averaging,bertsekas,bismarck}, BSP-style
gradient descent~\cite{mli}, and ADMM~\cite{boyd-admm}.

As we have discussed, our work builds on a trend towards asynchrony in
the learning literature. Among this literature, several studies stand
out. Most pragmatically, DimmWitted exploits the trade-offs inherent
in shared-memory statistical analytics~\cite{dimmwitted}; here, we
study the problem of multi-node coordination present in the
distributed environment, which brings higher latency and a different
set of trade-offs. This work is complementary to ours insofar as
our approach is largely agnostic to the local solver on each machine but
is instead concerned with coordinating between solvers.

More theoretically, Wei and Ozdaglar~\cite{wei-async-admm} propose a
variant of ADMM in which, in each round, a randomly chosen set of
processes (synchronously) takes an ADMM step. This algorithm and a
closely related mechanism proposed by Iutzeler et
al.~\cite{iutzeler-async-admm} provide excellent theoretical
convergence guarantees but are, nevertheless, synchronous, and are not
evaluated in practice~\cite{wei-async-admm}. Zhang et al. leverage
partial synchrony and bounded delay to similarly allow additional
asynchrony in ADMM execution~\cite{zhang-async-admm} and provide---as
one of few instances in the ADMM literature---an experimental
evaluation of their algorithm on a distributed 18-node MPI cluster and
synthetic datasets. Their bounded asynchrony is more constrained than our
\badmm implementation.

Several recent systems exploit a \textit{parameter server} to
facilitate state sharing during distributed asynchronous model
training~\cite{brain,parameterserver,cipar,adam}. These parameter
servers effectively act as a two-level aggregation tree for updates,
and individual parameter servers expose different data consistency
models, such as bounded staleness or causal consistency. These systems
closely resemble key-value stores with extensions for abstract data
types like vectors, do not support general computation (i.e., are
highly specialized for tasks like deep learning~\cite{brain,adam}),
and do not, in general, provide a complete cluster management
framework (i.e., the parameter server itself is used by a set of
parallel processes external to the server). Thus, while these systems
are indeed useful, we seek a more generally applicable architecture
for asynchronous state sharing that is compatible with existing,
widely deployed dataflow systems like Spark (i.e., one that does not
require installing another separate system simply for performing
complex analytics). Parameter servers simplify partial replication of
models, but, given the duality of message passing and shared
memory~\cite{duality1,abd}, such optimizations are also applicable to
the ASIP iterator interface.


\minihead{Related database systems concepts} Our resulting
\xchg implementation has several close relatives in the broader
database literature. As noted in Section~\ref{sec:algorithm}, standard
ADMM is reminiscent of the distributed numerical consensus achieved by
the Escrow method~\cite{escrow} and the Demarcation
protocol~\cite{demarcation}. Related techniques for re-balancing
and/or bounding numerical error across
replicas~\cite{olston-thesis,yu-efficient} are similarly applicable to
the problem of maintaining \badmm consensus variables, which we have
only touched upon in this work. Our implementation of \xchg is
reminiscent of Sideways Information Passing in traditional dataflow
architectures~\cite{ives-sideways}, which allows data transfer between
parallel operators and has been successfully applied in diverse
domains such as join~\cite{rstar} and magic set~\cite{magic-sip}
evaluation. We develop the \xchg iterator interface as a means of
achieving similar, fine-grained data sharing between a set of primary
operators that repeatedly accesses data (i.e., implement UDA
functionality for repeated passes of partitioned data). As a useful
lens on our techniques, the sharing of primal variables in \badmm can
be considered an instance of query shipping~\cite{kossmann-dqp}
(instead of data/delta shipping, as in~\cite{hogwild}).

Finally, the ASIP framework shares some similarities with the actor
model\cite{Hewitt73}. Indeed, (largely as a convenience) we leveraged
the Akka actor system implementation used in Spark for data transfer
in the ASIP iterator.  However, unlike actors, the ASIP iterators
are \emph{pull}-based, inherently data-centric, and their creation is
managed by the by dataflow system and not the user defined logic.
These restrictions simplify our design and provide additional latitude
in managing partitioning of the data and movement of information
across operators---providing a sweet spot in the design space between
a lower-level primitive like MPI~\cite{mpi} or sockets and higher-level
programming models such as actors.

%% file: conclusion.tex

\section{Conclusions and Future Work}
\label{sec:conclusion}

In this paper, we presented the ASIP abstraction to enable
asynchronous complex analytics within the context of traditional,
data-parallel (and otherwise synchronous) dataflow systems. ASIP
presents an iterator-centric programming model with a special
ASIP iterator used to communicate asynchronously between
concurrent iterator instances within a single dataflow stage. We
ported two popular convex programming algorithms---SGD and ADMM---to
our prototype implementation of the ASIP on top of the Spark dataflow system.
By leveraging the statistical robustness of these operators, we
provide fault tolerant implementations that substantially outperform
their synchronous counterparts written directly in Spark.

While the ASIP abstraction is targeted at asynchronous complex analytics tasks, it would be
interesting to extend them to more general computation tasks (e.g., early stopping
criteria for online aggregation~\cite{ola-partial}). This will undoubtedly require
modifications to the fault tolerance model, including more complex
checkpointing schemes.

We also believe it would be interesting to allow ASIP algorithms to be
executed deterministically in order to better support debugging and
replay. One strategy for accomplishing this would be to divide ASIP
operator execution into finite epochs (e.g., by periodically pausing
the operator execution thread) and only allowing ASIP operator
message delivery within epochs. Checkpointing each operator's input
queue (e.g., as an RDD) would enable deterministic replay, albeit at a
cost to both storage and runtime overheads.

Finally, we are interested in exploring alternative communication
patterns including aggregation trees to reduce the cost of ASIP
iterator use during the exchange of larger models than those we consider
here. We suspect that adapting mechanisms such as the demarcation
protocol~\cite{demarcation}, escrow transactional
method~\cite{escrow}, and approximate replication
techniques~\cite{olston-thesis} to track divergence between solvers
and reduce sharing will further reduce these costs.

%% file: appendix.tex
\normalsize
\linespread{1}

\appendix

\section{Experimental Details}
\label{sec:appendix-exp}

The choice of algorithm parameters can have a considerable impact in the overall performance of each algorithm.
We made a best effort to find consistent parameters settings for each of the algorithms.
While we explored tuning these parameters on a per-dataset basis we ultimately settled on a single set of consistently performing \emph{default} parameters which are summarized in \tableref{table:params}.

\begin{table}[h]
\begin{center}
\begin{tabular}{|l r|}
\hline
\multicolumn{2}{|c|}{\texttt{GD}} \\
Batch Size & all records in partition \\
SGD $\eta_0$ & 1e-1 \\
\hline\hline\multicolumn{2}{|c|}{\texttt{ADMM}} \\
Primal residual $\epsilon$ & 1.0e-5 \\
Lagrangian $\rho$ & 1.0e-2 \\
Primal solve max. SGD iterations & 10000 \\
SGD $\eta_0$ & 1e-1 \\
SGD batch size & 10 records\\
\hline\hline\multicolumn{2}{|c|}{\bsgd} \\
Maximum \xchg \texttt{push} rate & once per 10ms \\
SGD steps between \xchg \texttt{poll} requests & 10 \\
SGD $\eta_0$ & 1e-1 \\
SGD batch size & 10 records\\
\hline\hline\multicolumn{2}{|c|}{\badmm}\\
Maximum \xchg \texttt{push} rate & once per 100ms \\
Primal residual $\epsilon$ & 1.0e-5 \\
Primal solve max. SGD iterations & 10000 \\
Lagrangian $\rho$ & 1.0e-2 \\
Records per gradient step & 10 \\
SGD $\eta_0$ & 1e-1 \\
SGD batch size & 10 records\\
\hline
\end{tabular}
\end{center}
\caption{Summary of parameters used in experiments}
\label{table:params}
\end{table}

\subsection{Fault Tolerance Experiments:}

\begin{figure*}[ht]
\subfigure[forest]{
  \includegraphics[width=.23\linewidth]{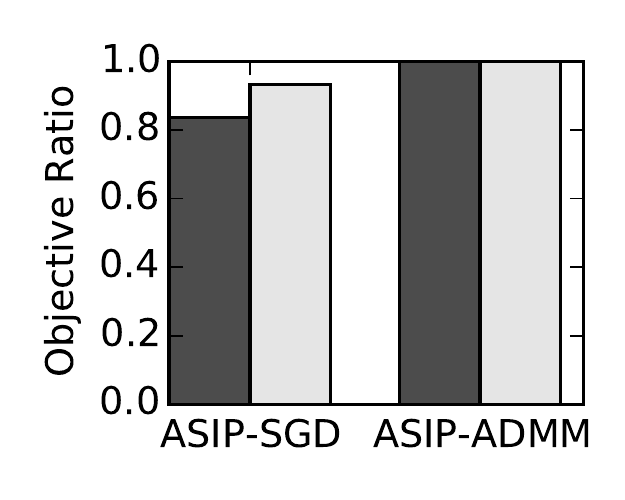}
}
\subfigure[flights]{
  \includegraphics[width=.23\linewidth]{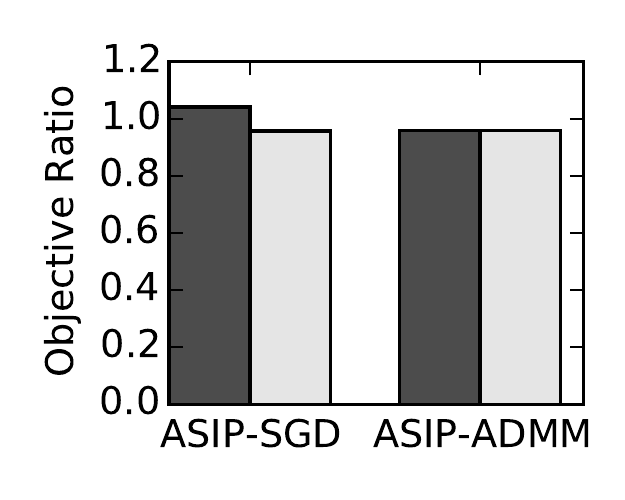}
}
\subfigure[dblp]{
  \includegraphics[width=.23\linewidth]{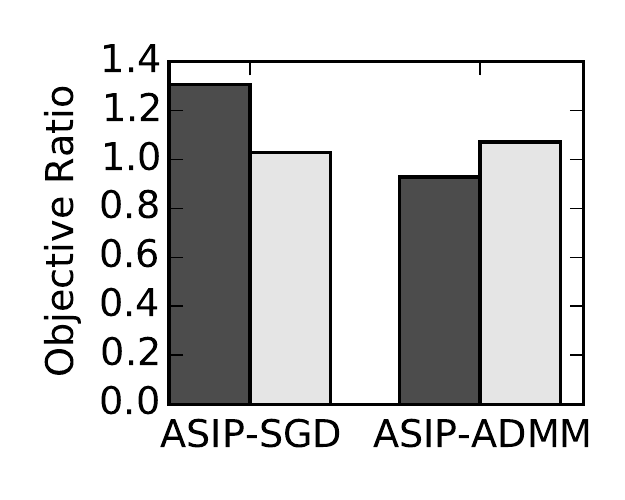}
}
\subfigure[wikipedia]{
  \includegraphics[width=.23\linewidth]{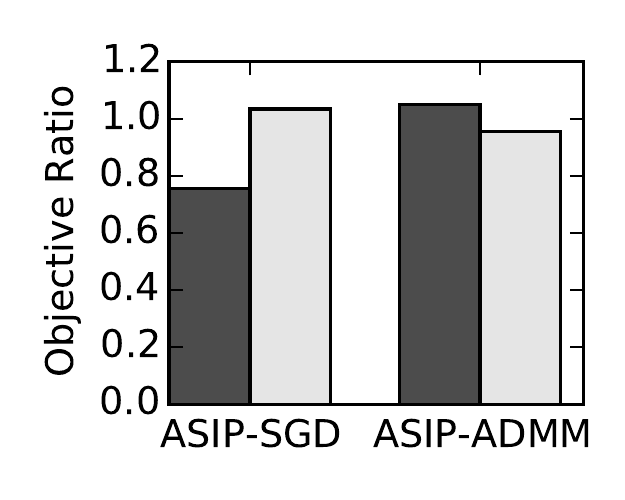}
}\vspace{-1em}
\caption{\footnotesize Additional Fault Tolerance Experiments. We plot the ratio of the objective (\texttt{SVM+L2}) after a machine reset 10 seconds into the computation and the objective without reset (lower is better).  The dark shaded bar is the objective ratio at 15 seconds and the light bar is at 30 seconds.  In general we find that the asynchronous algorithms are relatively robust to machine failure.}
\label{fig:faulttolerance-10seconds}
\end{figure*}

To better understand the fault tolerance behavior we also considered the effect of introducing a machine failure later in the program execution after convergence (at 10 seconds rather than 3 seconds).
In \figref{fig:faulttolerance-10seconds} we plot the ratio of the objective with a fault over the objective without a fault.
Again we observe that the algorithm naturally recovers to a similar objective value.

\section{ASIP Pseudocode}
\label{sec:appendix-pseudo}

In section 4 we provided a high-level sketch of the ASIP convex programming algorithms.
Here we provide a more detailed presentation of the implementation of these algorithms in the ASIP programming model using a scala like syntax as well as their simpler BSP counter-parts.
In both cases we note the similarity in their complexity and design.

In \listref{listing:ASIP-SGD} we implement the user defined function (UDF) for our distributed stochastic gradient descent solver.
At a high-level this algorithm closely follows the traditional serial algorithm with the added loop over the ASIP iterator and horizontal broadcast (\texttt{asip.push}).
In comparison, the basic batch gradient descent algorithm (\listref{listing:BSP-GD}) is not much simpler than the ASIP formulation of distributed asynchronous gradient descent.
The ASIP-DualAveraging algorithm (\listref{listing:ASIP-DualAveraging}) shares some similarity with the SGD algorithm though the primal updates are a more uniformly weighted sum of the dual updates.
In practice we found that the ASIP-SGD algorithm generally performs better.
\vspace{1em}

\begin{lstlisting}[language=scala, basicstyle=\ttfamily, columns=fullflexible,label=listing:ASIP-SGD, caption={\textbf{ASIP-SGD:} The implementation of the ASIP user defined function for the SGD algorithm}, mathescape]
def sgdUDF(data: Input,
           asip: ASIPIterator) = {
  // External Constants
  val eta = 1.0 // Learning rate
  val regParam = 1.0 // Regularization parameter

  var w = InitialModel()
  var wOld = null
  var t = 0
  for (t in 1 to T)
    wOld = w
    while (asip.hasNext) {
      w = w - (eta / sqrt(t)) * asip.next()
    }
    val (x, y) = data.nextWithLoop()
    val grad = lossGradient(w, (x, y)) +
      regParam * regGradient(w)
    w = w - (eta / sqrt(t)) * grad
    asip.push(grad)
  }
  return w
}
\end{lstlisting}

\begin{lstlisting}[language=scala, label=listing:BSP-GD, basicstyle=\ttfamily, columns=fullflexible, caption={\textbf{Batch Gradient Descent:} An implementation of batch gradient descent using the standard dataflow operators.}, mathescape]
// External constants
val eta = 1.0 // Learning rate
val regParam = 1.0 // Regularization parameter

var w = InitialModel()
var wOld = null
var t = 0
for (t in 1 to T) {
  wOld = w
  val grad = data.map { case (y, x) =>
    lossGradient(w, (x, y))
  }.avg() + regParam * regGradient(w)
  w = w - (eta / sqrt(t)) * grad
}
return w
\end{lstlisting}

\begin{lstlisting}[language=scala, basicstyle=\ttfamily, columns=fullflexible, label=listing:ASIP-DualAveraging, caption={\textbf{ASIP Dual-Averaging UDF}}, mathescape]
def dualAveragingUDF(data: Input,
    asip: ASIPIterator) = {
  // External constants
  val eta = 1.0 // Learning rate
  val regParam = 1.0 // Regularization parameter

  var dualSum = ZeroVector()
  var dual = ZeroVector()
  var w = InitialModel()
  var wOld = null
  for (t in 1 to T) {
    wOld = w
    while (asip.hasNext) {
      dualSum = dualSum + asip.next()
    }
    val (x, y) = data.nextWithLoop()
    val grad = lossGradient(w, (x, y)) +
      regParam * regGradient(w)
    dualOld = dual
    dual = dualSum / nWorkers + grad
    w = -(eta / sqrt(t)) * dual
    asip.push(dual - dualOld)
  }
  return w
}
\end{lstlisting}

In \listref{listing:ASIP-ADMM} we present the ASIP formulation of the ADMM algorithm.
The ASIP formulation, closely follows the synchronous variant of ADMM with a horizontal exchange stage before applying the consensus and dual updates.
In \listref{listing:BSP-ADMM} we present the similar BSP formulation.
As before the BSP and ASIP formulations share similar structure and implementation complexity.

\begin{lstlisting}[language=scala, basicstyle=\ttfamily, columns=fullflexible,label=listing:ASIP-ADMM, caption={\textbf{ASIP ADMM UDF}}, mathescape]
def admmUDF(data: Input,
    asip: ASIPIterator) = {
  // External constants
  val eta = 1.0 // Learning rate
  val regParam = 1.0 // Regularization parameter
  val rho = 1.0 // Consensus parameter
  val nodes = 128 // Number of iterators

  var primalAvg = ZeroVector()
  var dualAvg = ZeroVector()
  var consensus = InitialModel()
  var w = InitialModel()
  var wOld = Empty()
  var dual = ZeroVector()
  for (k in 1 to K) {
    wOld = w
    // Primal Update
    var t = 0
    while ( change(w) < eps ) {
      val (x, y) = data.nextWithLoop()
      val grad = lossGradient(w, (x, y))
        + dual + rho * (w - consensus)
      w = w - (eta / sqrt(t)) * grad
      t += 1
    }
    // Exchange
    asip.push((w - wOld, dual dualOld) / nodes)
    while (asip.hasNext) {
      (primalAvg, dualAvg) += asip.next()
    }
    // Consensus Update
    consensus = consensusProx(primalAvg, dualAvg)
    // Dual Update
    val dualOld = dual
    dual = dual + rho * (consensus - z)
  }
  return consensus
}
\end{lstlisting}

\begin{lstlisting}[language=scala, basicstyle=\ttfamily, columns=fullflexible, label=listing:BSP-ADMM,
caption={\textbf{BSP-ADMM:} A BSP implementation of ADMM using dataflow operators.}, mathescape]
// External constants
val eta = 1.0 // Learning rate
val regParam = 1.0 // Regularization parameter
val rho = 1.0 // Consensus parameter
val nodes = 128 // Number of iterators

var consensus = spark.Broadcast(InitialModel())
var consensusOld = Empty()
var primalDual = spark.parallelize(
  Array.fill(nodes)((InitialModel(), Zero()))
  )

for (k in 1 to K) {
  primalDual = data.zipPartitions(primalDual) {
    (data, Iterator(w), Iterator(dual)) =>
    // Primal Update
    var t = 0
    var w = InitialModel()
    val z = consensus.value
    while ( change(w) < eps ) {
      val (x, y) = data.nextWithLoop()
      val grad = lossGradient(w, (x, y))
        + dual + rho * (w - z)
      w = w - (eta / sqrt(t)) * grad
      t += 1
    }
    Iterator((w, dual))
    ).cache()

  // Collect primal and dual averages
  val (wAvg, dualAvg) = primalDual.avg()
  val z = consensusFun(wAvg, dualAvg)
  // Broadcast consensus value
  consensusOld = consensus
  consensus = spark.Broadcast(consensus)
  // Execute dual update in parallel
  primalDual.map { case (w, dual) =>
    dual + rho * (w - consensus.value)
  }
}
return consensus
\end{lstlisting}

\section{Mathematical Derivations:}
\label{sec:derivations}

The convex programming methods we considered here require that we be able to construct the gradient (or sub-gradient) of the loss and regularization terms in the objective.
Below we derive the corresponding sub-gradients for the loss functions:
\begin{align*}
\nabla_w \text{hingeLoss}(w, (y, x)) & = \nabla_w (1 - y w^T x)  \\
& = \begin{cases}
   -y w & \text{if } y w^T x < 1 \\
   0      & \text{otherwise }
  \end{cases} \\
\nabla_w \text{logisticLoss}(w, (y, x)) & = \\
\nabla_w (  (1-y) & \log \left(1 - \sigma(w^T x) \right) +  y \log \sigma(w^T x) )  \\
& = \left(y -\sigma(w^T x) \right)  x
\end{align*}
and for the regularization penalties:
\begin{align*}
\nabla_w L_2(w) & = \nabla_w \frac{1}{2}||w||_2^2  = w \\
\nabla_w L_1(w) & = \nabla_w |w|_1 = \begin{cases}
   -1 & \text{if } w < 0 \\
   [-1,1] & \text{if } w = 0 \\
   1 & \text{if } w > 0 \\
  \end{cases} \\
\end{align*}

\section{ADMM}

While Boyd et al.~\cite{boyd-admm} provide an exceptional overview of the mathematical background behind the ADMM algorithm, in this section we summarize the key details used in this work.

We employ the method of dual-decomposition to break the single convex programming problem into a collection of $p$ convex programming problems one for each of $p$ processors (iterators):
\begin{equation}
\begin{aligned}
& \underset{\text{wrt. } w_1, \ldots, w_p, z}{\text{minimize}}
& &  \frac{1}{|\data|}\sum_{i = 1}^p\sum_{r \in \data_i} \loss(w_i, r) + \lambda \reg(z)  \\
& \text{subject to}
& & w_i = z
\end{aligned}
\end{equation}
where we have partitioned the data across machines and constrained the solutions to the sub-problems to match a shared variable $z$.
Note that a solution to this problem is also a solution to the original convex programming problem.
We introduce an additional augmenting term to the above equation without changing the optimal values of $w_1, \ldots, w_p$ and $z$:
\begin{equation}
\begin{aligned}
& \underset{\text{wrt. } w_1, \ldots, w_p, z}{\text{minimize}}
& &  \frac{1}{|\data|}\sum_{i = 1}^p \left( \sum_{r \in \data_i} \loss(w_i, r) + \frac{\rho}{2}\TwoNorm{w_i - z}^2\right) + \lambda \reg(z)  \\
& \text{subject to}
& & w_i = z
\end{aligned}
\end{equation}
When the constraints are satisfied then $\frac{\rho}{2}\TwoNorm{w_i - z}^2 = 0$.
The introduction of this additional term will play an important role in smoothing the sub-problems and enabling an analytic $z$ update.

Thus far the constraints $w_i = z$ couple each sub-problem making it difficult to solve in parallel.
However, we can remove the constraints $w_i = z$ by introducing Lagrange multipliers and moving the constraints into the objective.
We thus obtain the following Lagrangian:
\begin{align}
& \mathcal{L}(\{w_i\}_1^p, z, \{\mu_i\}_1^p) = \nonumber \\
& \frac{1}{|\data|}\sum_{i = 1}^p \left( \sum_{r \in \data_i} \loss(w_i, r) + \mu_i^T (w_i - z) + \frac{\rho}{2}\TwoNorm{w_i - z}^2 \right) + \lambda \reg(z)
\end{align}
leading to the dual formulation of the problem:
\begin{align}
\max_{\mu_1, \ldots, \mu_p} \left( \min_{w_1,\ldots, w_p, z} \mathcal{L}(\{w_i\}_1^p, z, \{\mu_i\}_1^p) \right) \label{eqn:dualAppend}
\end{align}
Assuming strong duality (which can be shown under mild assumptions) the solution to this dual problem is a solution to our original problem.

We can solve the dual problem (\eqnref{eqn:dualAppend}) in stages by alternating between primal and dual updates, hence the name Alternating Direction Method of (dual) Multipliers.
The ADMM algorithm is then broken into the following sequence of iterates:
\begin{align}
  w_i & \leftarrow \arg\min_{w_i}  \left( \sum_{r \in \data_i} \loss(w_i, r) + \mu_i^T (w_i - z) + \frac{\rho}{2}\TwoNorm{w_i - z}^2 \right) \label{eqn:admmw}\\
  z & \leftarrow \arg\min_{z}  \sum_{i=1}^p \mu_i^T (w_i - z) + \frac{\rho}{2}\TwoNorm{w_i - z}^2  + \lambda \reg(z) \label{eqn:admmz} \\
  \mu_i & \leftarrow \mu_i + \rho (w_i - z) \label{eqn:admmmu}
\end{align}
The primal updates in equations \ref{eqn:admmw} and \ref{eqn:admmz}
solve  convex sub-problems while the dual update in equation
\ref{eqn:admmmu} adjusts the sub-problems towards agreement.

\balance

The ADMM decomposition of the convex programming problem has a few useful properties.
First \eqnref{eqn:admmw} and \eqnref{eqn:admmmu} can be solved locally on each processor without access to the data or state on either processors.
Second \eqnref{eqn:admmz} can be further simplified:
\begin{align}
z \leftarrow \argmin_{z} \, \lambda  \reg(z) + \frac{z^T p \rho}{2} \left(z - 2 \bar{w} - \bar{\mu}\right)
\end{align}
where $\bar{w} = (1/p)\sum_i w_i$ and $\bar{\mu} = (1/p) \sum_i \mu_i$ are the averages of the variables on each processor.